\documentclass{article}

\usepackage{url}

\usepackage{amsmath}
\usepackage{graphicx}
\usepackage{pdfpages}
\usepackage{ifpdf}
\ifpdf
  \usepackage{epstopdf}
\fi
\usepackage[plain,noend]{algorithm2e}
\usepackage{color}
\usepackage{float}
\usepackage{subfigure} 
\usepackage{bbm}
\def\bSig\mathbf{\Sigma}

\newcommand{\latin}[1]{{\it #1}}
\usepackage{amsmath}
\usepackage{lipsum,times,graphicx,hyperref,cleveref}
\newtheorem{theorem}{Theorem}
\newtheorem{corollary}{Corollary}
\newtheorem{condition}{Condition}
\usepackage{enumitem}
\newlist{steps}{enumerate}{1}
\setlist[steps, 1]{label = Step \arabic*:}

\usepackage{PRIMEarxiv}

\usepackage[utf8]{inputenc} 
\usepackage[T1]{fontenc}    
\usepackage{hyperref}       

\usepackage{booktabs}       
\usepackage{amsfonts}       
\usepackage{nicefrac}       
\usepackage{microtype}      
\usepackage{lipsum}
\usepackage{fancyhdr}       
\usepackage{graphicx}       
\graphicspath{{media/}}     

\title{High-dimensional Feature Screening for Nonlinear Associations With Survival Outcome Using Restricted Mean Survival Time
\thanks{\textit{\underline{Citation}}: 
\textbf{Authors. Title. Pages.... DOI:000000/11111.}} 
}

\author{
  Y. Chen \\
  Department of Statistics and Actuarial Science \\
  The University of Hong Kong \\
  Hong Kong\\
   \And
   K.F. Lam \\
  Department of Statistics and Actuarial Science, \\Centre for Quantitative Medicine \\
  The University of Hong Kong,\\ Duke-NUS Medical School \\
  Hong Kong, Singapore\\
  \texttt{hrntlkf@hku.hk} \\
  \And
  Z. Liu \\
  Department of Biostatistics\\
  Columbia University \\
  New York, NY, U.S.A.\\
}

\begin{document}
\maketitle

\begin{abstract}
Feature screening is an important tool in analyzing ultrahigh-dimensional data, particularly in the field of Omics and oncology studies. However, most attention has been focused on identifying features that have a linear or monotonic impact on the response variable. Detecting a sparse set of variables that have a nonlinear or non-monotonic relationship with the response variable is still a challenging task. To fill the gap, this paper proposed a robust model-free screening approach for right-censored survival data by providing a new perspective of quantifying the covariate effect on the restricted mean survival time, rather than the routinely used hazard function. The proposed measure, based on the difference between the restricted mean survival time of covariate-stratified and overall data, is able to identify comprehensive types of associations including linear, nonlinear, non-monotone, and even local dependencies like change points. This approach is highly interpretable and flexible without any distribution assumption. The sure screening property is established and an iterative screening procedure is developed to address multicollinearity between high-dimensional covariates. Simulation studies are carried out to demonstrate the superiority of the proposed method in selecting important features with a complex association with the response variable. The potential of applying the proposed method to handle interval-censored failure time data has also been explored in simulations, and the results have been promising. The method is applied to a breast cancer dataset to identify potential prognostic factors, which reveals potential associations between breast cancer and lymphoma.
\end{abstract}


\section{Introduction}
\label{s:intro}
In recent years, rapid advances in Omics technologies have led to an increased biological understanding of cancer through a large array of ultrahigh-dimensional complex data, including gene expression. To relate these ultrahigh-dimensional features to a time-to-event outcome, feature screening needs to be performed to identify a sparse set of important variables before a refined analysis \cite{fu2018}. However, feature screening for survival data remains a challenge in detecting a set of variables that are related to the response variable in a nonlinear or non-monotone way. In fact, modern biomedical research has pointed out that besides linear effects, nonlinear and non-monotonic association between the expression levels of genes or the demographic features and survival outcomes also play an important role in large scale biological datasets. An example in breast cancer study is the effect of age at diagnosis on the progression free survival of breast cancer, where patients’ risk of death decreases with age until menopausal period and then increases \cite{admin1986, lee2020}. Another example is a lower or higher expression level of EGFR gene is found to be associated with poor survival of breast cancer patients \cite{kreike2010}.

Many \latin{ad hoc} studies advocated the Cox proportional hazards (PH) regression-based screening approach \cite{tibshirani2009,fan2002,zhao2012,yang2016,yi2022}. These methods work well in identifying a sparse set of variables that are linearly related to the response variable, but are generally inadequate when the relationship is not linear. \cite{wu2020} proposed a partially linear additive Cox model-based penalization procedure to accommodate nonlinear associations. However, it will be infeasible for ultrahigh dimensional data with a lack of stability due to the dramatic increase of dimensionality brought by using splines. Some of the model-free screening methods are able to identify the important features with a nonlinear association with survival endpoints, but restricted to monotone relationship. For example, \cite{song2014} proposed a model-free censored rank independence screening called CRIS based on inverse probability-of-censoring weighted rank correlation. \cite{zhang2017} developed a novel correlation rank based screening called CRSIS, which is built upon rank correlation between covariate and survival function. Other methods include a robust screening based on Pearson’s correlation of the covariate distribution and survival function \cite{hao2019}, a Spearman correlation screening procedure \cite{yan2017}, a Gehan-type rank screening accommodating heterogeneous censoring \cite{xu2020}, and a network-adjusted Kendall’s tau measure for feature screening \cite{wang2021}.  While correlation coefficients are useful for capturing monotonic associations, their ability to identify more complex nonlinear associations is limited. Non-monotonic associations, such as quadratic relationships or local dependencies like change points, may be overlooked by these coefficients.

Based on the aforementioned reasons, we present in this article a model-free screening device for right-censored data by introducing a novel robust measure built upon restricted mean survival time (RMST). The proposed measure captures the impact of covariates on mean life time in a comprehensive manner, including linear, nonlinear, local dependence, non-monotonic and piece-wise relationships, making it more flexible than most existing methods. This is achieved by integrating the absolute difference between covariate-stratified RMST and overall RMST over the range of the covariate. Notably, the proposed measure is parsimonious and estimation is simple as no tuning parameters are required. In addition, our approach can be extended to an iterative screening procedure with follow-up variable selection using the Cox additive model. This iterative screening method addresses issues associated with the curse of dimensionality, and multicollinearity among features. Comparing with the recent SII screening \cite{li2016}, which shows good performance in non-linearity detection, the proposed method is more robust in its design, which will be shown in the simulation studies. We demonstrate empirically that the proposed marginal screening method is highly adaptable and can be applied effectively to interval-censored data. 

Besides the afore-mentioned merits, the proposed feature screening method also advances the toolbox of routine marginal screening methods by providing a brand-new perspective in evaluating the impact of covariates on survival. Rather than building the measure of importance upon hazard ratio (HR) function as in the literature, this is the first attempt to formulate feature screening from the perspective of RMST. Merits are in multiple ways. First,  the mean survival time over a given period has a more intuitive and clinically relevant interpretation than hazard ratios \cite{andersen2004,chen2001}. As a result, the selected set of variables is more practical and easier to interpret for further analysis.  Second, traditional HR-based screening methods \cite{yi2022, yang2016} have a high demand for the Cox model or PH assumptions, which can be unreliable when these assumptions are not met. In contrast, the proposed method, which is motivated by the increasing recognition of RMST as an alternative to HR \cite{royston2013, huang2018, eaton2020}, can measure the covariate effect in a comprehensive way with no distribution assumption. This property and flexibility is particularly important in the context of immunotherapy development \cite{huang2018}.

The remainder of the paper is organized as follows. Section 2 motivates the use of expected RMST difference in feature screening for right-censored data, and presents the asymptotic distribution of the difference in sample RMSTs. Section 3 outlines the proposed screening procedure and its theoretical properties, along with an iterative screening algorithm that deals with multicollinearity. In Section 4, we report the results of large-scale simulation studies, while in Section 5, we analyze a breast cancer example. Finally, Section 6 concludes the paper with a summary and discussion of the findings. All proofs are provided in the appendix.

\section{RMST Difference Measure}
Consider a failure time study consisting of a random sample of $n$ individual subjects and $T$ being the corresponding failure time of interest. Denote $C$ be the censoring time, let $Y=\min\{T,C\}$ be the observed event time and $\Delta=I(T\leq C)$ be the censoring indicator. 
Moreover, let $\boldsymbol{X}=(X_1,\cdots,X_{p_n})^T$ be a $p_n$-dimensional vector of covariates, such as biomarkers or gene expressions. Then the observed data is denoted as $W_i=(\textbf{X}^T_i,Y_i,\Delta_i), i=1,\cdots,n$
, where $W_i$'s are identically and independently distributed. Throughout the paper, we assume an independent censoring mechanism.
	
Let the survival function be $S(\cdot)$. The restricted mean survival time (RMST), defined as the mean survival time restricted by any specified time point $\tau>0$ with $P(Y>\tau)>0$, is given by
	\begin{equation}
		\label{e:rmst}
  RMST(\tau)=E(T\wedge\tau)=\int_0^{\tau}S(t)dt.	
	\end{equation}
RMST is often used to evaluate the difference of survival time between treatment groups when the proportional hazards assumption is not justified, such as crossings in survival curves, short-term treatment effects and so on.

To investigate the relationship between a covariate $X_j$ and the survival outcome, we can use the concept of RMST. Define the two-halves covariate-stratified restricted mean survival time by 
	$$RMST_j^1(\tau^1_j;x)=\int_0^{\tau^1_j}S(t|X_j\geq x)dt$$ and $$RMST_j^2(\tau^2_j;x)=\int_0^{\tau^2_j}S(t|X_j< x)dt,$$
where $x$ can be any real number within the domain of the covariate $X_j$, and $\tau^1_j$ and $\tau^2_j$ are the specified restriction times such that $P(Y>\tau^1_j|X_j\geq x)>0$ and $P(Y>\tau^2_j|X_j< x)>0$. We can compare the covariate-stratified RMSTs with the overall $RMST(\cdot)$ to quantify the covariate effect on RMST as $|RMST_j^1(\tau^1_j;x)-RMST(\tau^1_j)|$ and $|RMST_j^2(\tau^2_j;x)-RMST(\tau^2_j)|$. 
Intuitively, a larger difference between the covariate-stratified RMST and overall RMST indicates a larger changes in mean life time after stratifying $X_j$ by $x$, which suggests a potential association between the covariate and the response. To capture the overall discrepancy of covariate-stratified RMST and overall RMST across all possible values of $X_j$, we propose the following global RMST discrepancy for the $j$th covariate:
	\begin{equation}
	\begin{split}
	     d_j=d_{j1}+d_{j2}=&\int_{\mathcal{X}_j}|RMST_j^1(\tau^1_j;x)-RMST(\tau^1_j)|dF_j(x)+\\
	     &\int_{\mathcal{X}_j}|RMST_j^2(\tau^2_j;x)-RMST(\tau^2_j)|dF_j(x),
	\end{split}
	\end{equation}
    where $F_j(x)$ is the cumulative distribution function of $X_j$. 
	This measure captures all the absolute RMST differences between the stratified group $X_j< x$ as well as $X_j\geq x, x\in\mathcal{X}_j$ and the unstratified sample across all possibilities of $x$. 
 Consequently, if $X_j$ has significant effects on survival outcome $T$, the global RMST discrepancy over the whole range of $X_j$ should increase. On the other hand, when $X_j$ is independent of the survival outcome $T$, $d_j$ should be 0 since the restricted samples from neither $X_j< x$ nor $X_j\geq x$ will show a different mean survival time from the overall RMST. Noted that we add up $d_{j1}$ and $d_{j2}$ instead of only considering one of them so that more information of local dependency will be incorporated in this measure, particularly when the local dependency is at the extreme tail part.
	
Several methods for the estimation of RMST are available in the literature such as pseudo-observations \cite{andersen2004}, M-estimation \cite{wang1999} and so on. In this article, we adopt the most dominant one, namely the Kaplan-Meier (KM) method \cite{meier1975}, as it provides an approximately unbiased estimate to RMST and is asymptotically normally distributed \cite{meier1975}. Specifically, for a given time point $0<\tau<\sup_{t\in \mathcal{T}}\{t\}$, we have
	$$\widehat{RMST}(\tau)=\int_0^\tau\hat{S}(t)dt,$$
	where $\hat{S}(t)=\prod_{i: Y_i\leq t}\big[1-\frac{\Delta_i}{R_i}\big]$ is the KM estimator of the survival function, and $R_i$ is the size of the risk set right before $Y_i$. To estimate the proposed measure $d_j$, one should be careful when setting the restriction time $\tau$, as the non-parametric Kaplan-Meier survival curve is not estimable after the largest observed failure time. As suggested by \cite{zhao2016}, using the largest uncensored observation as estimate for $\tau$ can maintain the asymptotic normality of RMST estimator and make use of the survival information as much as possible. Hence we set $\hat{\tau}^1_{ji}=\max_{\{k:X_{jk}\geq X_{ji},\ \Delta_k=1\}}\{Y_k\}$ and $\hat{\tau}^2_{ji}=\max_{\{k:X_{jk}< X_{ji},\ \Delta_k=1\}}\{Y_k\}$ for $i=1,\cdots,n$. Then the estimation of $d_j$ can be easily obtained by taking the sample average of the difference of the estimated RMST, i.e.,
$$\hat{d}_j=\frac{1}{n}\sum_{i=1}^n\bigg[|\widehat{RMST}_j^1(\hat{\tau}^1_{ji};X_{ji})-\widehat{RMST}(\hat{\tau}^1_{ji})|+|\widehat{RMST}_j^2(\hat{\tau}^2_{ji};X_{ji})-\widehat{RMST}(\hat{\tau}^2_{ji})|\bigg].$$
$\widehat{RMST}_j^1(\hat{\tau}^1_{ji};X_{ji})$ and $\widehat{RMST}_j^2(\hat{\tau}^2_{ji};X_{ji})$ are estimated based on $\hat{S}(t|X_j\geq X_{ji})$ and $\hat{S}(t|X_j< X_{ji})$, respectively. In fact, the proposed estimator $\hat{d}_j$ is invariant to a monotone transformation of the covariates, whether it is known or unknown. We consider only the $\widehat{RMST}$ estimated from the restricted sample group with a size greater than 5 to make a valid estimation of RMST and avoid any effect from outliers. 
	
 To theoretically illustrate the validity of the proposed measure, we examine the hypothesis test $H_{0j}:d_{j1}=0$ versus $H_{1j}:d_{j1}>0$. The following theorem shows that the proposed estimator $\hat{d}_{j1}$ has the same large sample property as $\hat{d}_{j2}$, which is the same as that of $\hat{d}_{j2}$ even though they are not independent.
    \begin{theorem}Under $H_{0j}$, we have 
		$$\sqrt{n}\hat{d}_{j1}\stackrel{\mathrm{w}^*}{\to}\int_{\mathcal{X}}\big|\int_{\mathcal{T}} \mathbbm{G}_j(t,x) dt\big|dF_j(x),$$
		where $\mathbbm{G}_j(t,x)$ is a Brownian bridge with mean 0 and $\stackrel{\mathrm{w}^*}{\to}$ means converge weakly.
		
		Under $H_{1j}$, suppose the Lebesgue measure of $\Omega(\kappa_j)=\{(t,x):|\int_\mathcal{T}[S_j(t;x)-S(t)]dt|<\kappa_j,(t,x)\in\mathcal{T}\times\mathcal{X}\}$ is 0, where $S_j(t;x)=P(T>t|X_j\geq x)$, then we have 
		$$\sqrt{n}(\hat{d}_{j1}-d_{j1})
		\stackrel{\mathrm{w}^*}{\to} \int_{\mathcal{X}}sgn\bigg(RMST_{j}^1(\tau_j^{1})-RMST(\tau_j^{1})\bigg)\int_{\mathcal{T}}\mathbbm{G}_j(t,x)dtdF_j(x).$$
	\end{theorem}
The above theorem establishes the asymptotic distributions of $\hat{d}_{j1}$ under the null and the alternative hypotheses. The proof is provided in Web Appendix A, and it further validates the proposed statistic by demonstrating that there is a discernible pattern for separating noise from signals. A similar result can be obtained for $\hat{d}_{j2}$. Of course, one can only use $\hat{d}_{j1}$ or $\hat{d}_{j2}$ as a measure of dependency between $X_j$ and the response, but they only concentrate on one direction of the stratification in the covariate thus induces information lost. While there is no theoretical finding on the asymptotic distribution of $\hat{d}_j$ due to the dependence between $\hat{d}_{j1}$ and $\hat{d}_{j2}$, we empirically demonstrate the superiority of $d_j$ over $d_{j1}$ or $d_{j2}$ using a simulation example in Web Appendix B. In that example, 5 different types of associations are considered and it turns out that $\hat{d}_j$ has a better performance on the whole. 

    Notice that the proposed method is applicable to right-censored data, but as demonstrated in \cite{zhang2020}, RMST is still estimable for interval-censored data. In this article, we also investigate the extension of the proposed measure to interval-censored response in an empirical manner and leave the theoretical work for future studies. For interval-censored data, we employed the self-consistency algorithm proposed by \cite{turnbull1976} to estimate the survival function $\hat{S}(t)$ and linear smoothing to estimate RMST, as outlined in \cite{zhang2020}. \cite{zhang2020} also demonstrated that the RMST estimator is asymptotically normal for interval-censored data. 
    
	A special case is when there is no censoring, and the response variable is a continuous variable. Suppose we have a random sample
 $\{Y_i,X_i\equiv(X_{i1},\cdots,X_{ip_n})^T\},i=1,\cdots,n$ from the population $\{Y,X\equiv(X_{1},\cdots,X_{p_n})^T\}$, where $Y$ is the response variable. The measure can be derived by
	\begin{equation*}
		\begin{split}
		d_j = & d_{j1}+d_{j2} =E|E(Y|X_j\geq x)-E(Y)|+E|E(Y|X_j< x)-E(Y)|\\
		& =\int\bigg|\int yf(y|X_j\geq x)dy-\int yf(y)dy\bigg|dF_j(x)+\\
		&\ \int\bigg|\int yf(y|X_j< x)dy-\int yf(y)dy\bigg|dF_j(x),\\
		\end{split}
	\end{equation*}
	where $f(y)$ is the density of random variable $Y$ and $f(y|X_j\geq x)$ is the conditional density of $Y$ given $X_j\geq x$.
	Then the estimator of $d_j$ takes the form
	\begin{equation}
        \label{e:complete}
		\begin{split}
			\hat{d}_j =  \hat{d}_{j1}+\hat{d}_{j2} =& \frac{1}{n}\sum_{k=1}^n\bigg|\frac{1}{\#\{i:X_{ji}\geq X_{jk}\}}\sum_{i=1}^nY_iI(X_{ji}\geq X_{jk})-\frac{1}{n}\sum_{i=1}^nY_i\bigg|+\\
			& \frac{1}{n}\sum_{k=1}^n\bigg|\frac{1}{\#\{i:X_{ji}< X_{jk}\}}\sum_{i=1}^nY_iI(X_{ji}< X_{jk})-\frac{1}{n}\sum_{i=1}^nY_i\bigg|.
		\end{split}
	\end{equation}
   This measure can also be applied in the screening procedure for uncensored data and we will apply it in the iterative screening procedure proposed in Section~\ref{ss:iter}.
	
\section{Sure Independence Screening Using RMST-based Index}
\subsection{Sure Independece Screening}
For ultrahigh-dimensional data, the number of variables $p_n$ increases exponentially with sample size $n$. Therefore we propose a model-free screening procedure based on the proposed $\hat{d}_j$ to identify the set of active variables denoted by 
	
	\centerline{$\mathcal{M}_*=\{j:E(T|\textbf{X})$ functionally depends on some $\textbf{X}_j\},$}
	
\noindent with a non-sparsity size $s_n=|\mathcal{M}_*|$. Noted that `functionally' means it captures all types of dependencies, whether they are local or global, linear or nonlinear, monotonic or non-monotonic. Utilizing the proposed $\hat{d}_j$, we select the following set with a prespecified threshold $\gamma_n$,
	$$\hat{\mathcal{M}}=\{j:\ \hat{d}_j\geq \gamma_n,\ j=1,\cdots,p_n\},$$
	as the estimate for $\mathcal{M}_*$. In practice, we often choose $\gamma_n$ such that the top $[n/\log n]$ markers are remained in the selected set. As the recent attention has drawn to the false discovery rate control method such as knockoff \cite{barber2015}, we propose that the size of selected set can be determined by the threshold of knockoff under a target false discovery rate $\alpha$, similar to the work by \cite{liu2022}. Specifically, knockoff can be done based on the important set selected with suree independence screening property. Then, it can be proved that false discovery rate control and sure independence screening properties hold simultaneously. However, this is beyond the scope of this article and readers can refer to \cite{liu2022} for more details.
	
	Next, we are going to study the theoretical properties inherited by the proposed screening procedure. The following conditions are required to establish the properties:

\begin{condition}
\label{con1}
    Let $S_{X_j}(\cdot)$ and $S_C(\cdot)$ be the marginal survival function of $X_j$ and $C$, respectively. Suppose that given $t\in\mathcal{T},x\in\mathcal{X}_j,$ there exists constants $\gamma,\eta$, and $\lambda$, such that $0<\gamma<S(t_0)S_C(t_0),0<\eta\leq S_{X_j}(x)$ and $0<\lambda\leq Pr(T>t_0,X_j>x)$ for $1\leq j \leq p_n$.
\end{condition}
\begin{condition}
\label{con2}
    For some positive constant $m$, $\lim_{p_n\to\infty}(\min_{j\in\mathcal{M}_*}d_j-\max_{j\in\bar{\mathcal{M}}_*}d_j)>m$.
\end{condition}
\begin{condition}
\label{con3}
    $\min_{j\in\mathcal{M}_*}d_j\geq c_0n^{-\phi}$ for some $0<\phi<1/2$ and $c_0>0$.
\end{condition}

	Condition~\ref{con1} guarantees a robust estimation of the impact of $X_j$ on the distribution of $T$ given the information in $\mathcal{T}\times\mathcal{X}_j$. Condition~\ref{con2} assumes that there is a difference between the signal and the noise in the statistic and is necessary to establish the ranking consistency property. Condition~\ref{con3} ensures that the order of the statistic $\hat{d}_j$ from the active set is at least $n^{-\phi}$ and will not be too small to find out. This condition guarantees the sure screening property without imposing any specific model assumption on the data. Then, we show that the following theorem establishes the sure screening property. The proof can be found in Web Appendix A.
	
	\begin{theorem}
 \label{th:ss}
		 Under Conditions 1,2, and 3, setting $\gamma_n=\frac{1}{2}c_0n^{-\phi}$ leads to the following sure screening property
		$$Pr\bigg(\mathcal{M}*\subset\hat{\mathcal{M}}\bigg)\geq1-s_nb_1\exp\{a_n\},$$
		where $b_1$ is a constant, $s_n$ is the size of the active set and $a_n$ goes to negative infinity as $n$ goes to infinity.
	\end{theorem}

	\begin{corollary}[Controlling False Discoveries]
 \label{co:fdr}
		Under Conditions 1 and 3, we have 
		$$Pr\bigg(|\hat{\mathcal{M}}|\leq \frac{4}{c_0}n^{\phi}\sum_jd_j\bigg)\geq 1-p_nb_1\exp\{c_n\},$$
  and $c_n$ goes to infinity as $n$ goes to infinity.
	\end{corollary}
	
Theorem \ref{th:ss} states that the proposed screening procedure ensures the sure screening property, i.e., it can identify all the variables in the true active set with probability 1 even when $s_n=o(\log n)$ and $p_n=o(\exp{n})$ as $n$ increases to infinity. The order of $p_n$ is higher than most existing feature screening methods, which typically assume $p_n=o(\exp{n^{1-2\phi}})$, so that higher dimensionality can be tolerated in the proposed method. Notably, the proposed method is more flexible and robust than existing methods since it does not require any conditions on the tail probability or the underlying models, making it suitable for heavy-tailed covariates with contamination and unknown model assumptions. Corollary \ref{co:fdr} shows that the size of the selected set can be controlled at a polynomial level, and so can the size of the false discoveries.

\subsection{Iterative Screening Procedure}
\label{ss:iter}
The proposed screening procedure has two limitations that need to be addressed. Firstly, there is no standard way to determine the number of variables selected. Secondly, the use of marginal utility in the screening process may overlook predictors that are marginally irrelevant but jointly highly associated with the response. This issue is exacerbated by multicollinearity, where only some of the true predictors may be selected, while their correlated counterparts are ignored. To overcome these challenges, we propose an iterative version of the marginal screening procedure in this subsection. The size of the selected set can be determined in a data-driven way. The iterative algorithm enhances the power of identifying the true biomarkers, as demonstrated in the final example in the simulation studies.

The iterative procedure is summarized in Algorithm 1. Motivated by the Iterative Sure Independence Screening (ISIS) proposed by \cite{fan2008}, we repeatedly utilize information in the deviance residuals \cite{davison1989} of the model built in the previous step. To maintain its capability of detecting nonlinear association, we adopt an additive Cox model with Lasso variable selection in each iteration. The procedure proceeds in a data-driven way, and it continues until no more variables are selected by the model. \\

\begin{algorithm}
Iterative Screening Procedure.
\begin{tabbing}
\qquad \qquad \KwIn{$\ q, (\textbf{X}^T_i,Y_i,\Delta_i),i=1,\cdots,n.$}\\
\qquad \qquad \textbf{Initialize:} 
$k=1,\ \mathcal{A}^1=\emptyset.$
Apply the RMST screening procedure and select the top\\ \qquad \qquad $[n/\log n]$ variables as the candidate set $\mathcal{A}^0$. \\
\qquad \qquad \textbf{while} $|\mathcal{A}^k|<q$ and $\mathcal{A}^k\neq\mathcal{A}^{k-1}$\\
     \qquad \qquad \qquad 1. Conduct variable selection among $\mathcal{A}^{k-1}$ via Lasso penalty in the generalized\\ \qquad \qquad \qquad additive Cox model and obtain the selected set $\mathcal{A};$ \\
     \qquad \qquad \qquad 2. Fit a generalized additive Cox model on $\mathcal{A}$ and obtain the deviance residual for\\ \qquad \qquad \qquad each sample. Treat the deviance residuals as new response and conduct screening\\ \qquad \qquad \qquad  procedure using the criterion in Equation (\ref{e:complete}) for uncensored data on covariate set\\ \qquad \qquad \qquad $\{1,\cdots,p_n\}\backslash \mathcal{A}$ with top $[n/\log n]$ selected. Denote the selected set as $\mathcal{M};$\\
     \qquad \qquad \qquad 3. $k=k+1$ and $\mathcal{A}^{k}=\mathcal{A}\cup\mathcal{M}$.\\
     \qquad \qquad \KwOut{$\mathcal{A}^k.$}\\
   \end{tabbing}
\end{algorithm}

Notice that $q$ is the final size of the selected set and usually $q<n$ such as $q= [n/\log n]$ in the simulation studies. In fact, any marginal screening procedure can be adopted in the initialization procedure, and we will compare different initialization methods in Scenario 5 of the simulation studies.
In particular, the additive Cox model in Step 1 is built based on the hazard function
$$\lambda(t|\boldsymbol{X}_{\mathcal{A}})=\lambda_0(t)\exp\{\sum_{j=1}^{|\mathcal{A}|}\Psi_j(X_j)\},$$
where $\lambda_0(t)>0$ is an arbitrary baseline hazard function and $\Psi_j(X_j),j=1,\cdots,|\mathcal{A}|$ are some unknown functions of the variable $X_j$. We propose to approximate them using B-splines, i.e., 
	$$\Psi_j(X_j)=\sum_{i=1}^{m_j}\alpha_{ji}N_{ik}(X_j), j=1,\cdots,|\mathcal{A}|,$$
	where $\{N_{ik}\}_{i=1}^{m_j}$ are B-splines with degree $k$ and $m_j$ is the number of knots. In practice, we suggest $k=3$ corresponding to the cubic spline. $\boldsymbol{\alpha}=(\alpha_{11},\cdots,\alpha_{1m_1},\cdots,\alpha_{|\mathcal{A}|1},\cdots,\alpha_{|\mathcal{A}|m_{|\mathcal{A}|}})'$ is the vector of unknown parameters. Further variable selection via lasso is conducted by maximizing the penalized log partial likelihood 
	$$l(\boldsymbol{\alpha})=\sum_{i=1}^n\delta_i\bigg[\sum_{j=1}^{|\mathcal{A}|}\Psi_j(X_{ji})-\log\sum_{l\in\mathcal{R}(Y_i)}\exp\big\{\sum_{j=1}^{|\mathcal{A}|}\Psi_j(X_{jl})\big\}\bigg]-\theta\sum_{j=1}^{|\mathcal{A}|}\sum_{i=1}^{m_j}|\alpha_{ji}|,$$
	where $\mathcal{R}(Y_i)=\{l:Y_l\geq Y_i\}$ is the risk set and $\theta$ is the non-negative shrinkage parameter. We adopt the cross-validated log-likelihood function ($cvl$) to choose $\theta$ such that $cvl$ is maximized following the work by \cite{verweij1993}. We set the selected set in each iteration as $\mathcal{A}^*=\{j:\exists\ k\in m_j, s.t.\  \hat{\alpha}_{jk}\neq0,\ j=1,\cdots,|\mathcal{A}|\}.$  The above procedure can be implemented using the 'splines' and 'biospear' package in R and it turns out that a cubic spline with $k=3$ and $m_j=3$ is good enough in general.

\section{Simulation Studies}
	In this section, we investigate the finite sample performance of the proposed method under right-censored data by evaluating it in 4 scenarios. We compare the proposed RMST screening with four existing methods including the model-based Lasso by \cite{tibshirani2009}, the model-free correlation rank sure independence screening (CRSIS) by \cite{zhang2020}, censored rank independence screening (CRIS) by \cite{song2014}, and survival impact index screening (SII) by \cite{li2016}. We also include an additional scenario to evaluate the performance of our proposed iterative RMST screening algorithm in the presence of strong collinearity with right-censored data. These five scenarios include the linear transformation model, additive Cox model, and a piece-wise model with change-points in covariates. Additionally, although the theoretical work for interval-censored data has not yet been developed, we employ the screening procedure built upon the linear smoothing of RMST described in Section 2 \cite{zhang2020} to evaluate the empirical performance for interval-censored data and compare it with that of \cite{hu2020}.
	
	Three criteria are employed to evaluate the performance of feature screening, namely, Median, IQR, and $P_{all}$. Median and IQR represent the median number and the interquartile range of the smallest model size that can incorporate all true active predictors when the model size is not fixed. Out of 100 replications, $P_{all}$ is the proportion of all active predictors selected for a fixed model size $[n/\log n]$ out of 100 replications. A more effective method will have $P_{all}$ closer to 1 and smaller Median and IQR values.
		
	\textbf{Scenario 1.} We first consider the following linear transformation model adopted from \cite{song2014}:
	$$H(T)= - \beta^T X+\epsilon,$$
	where $H(t)=\log\{\frac{1}{2}(e^{2t}-1)\},\ \beta=(1,0.9,\textbf{0}_6,-0.8,-1)^T$, $X=(X_1,\cdots,X_p)^T$ is a $p\ $-dimensional vector of covariates following a multivariate normal distribution with mean zero and covariance matrix $\Sigma$. The elements of $\Sigma$ have a first-order autoregressive structure, with $\Sigma{ij}=0.5^{|i-j|}$. We consider three error distributions: the standard normal distribution, standard extreme distribution, and standard logistic distribution correspond to normal transformation model, proportional hazards model, and proportional odds model, respectively. We generate censoring times from a uniform distribution $[0,u]$ such that the average censoring rate is around $20\%$ and $40\%$. We consider sample sizes of $n=200$ and $n=100$ and a dimension of covariates of $p=2000$. The model size is chosen to be $[n/\log{n}]$.
 
	The simulation results are shown in Table~\ref{tab:Table1}. We find that when $n=100$ the proposed RMST screening has comparable performance to CRIS and CRSIS, which are known to work well for linear effects. When $n$ increases to 200, all methods show sure screening property with $P_{all}$ reaching 100\%. Among different error distributions, the screening performance under the PH model is the best. Additionally, the proposed method is less sensitive to changes in censoring rate compared to other methods and can handle low event rates.

	\textbf{Scenario 2.} To examine the screening performance in the non-linear and non-monotone relationship scenario, we consider the following additive model
	$$H(T)=-4g_1(X_1) + 2.2g_2(X_2)-
	2.4g_3(X_3)+0.9g_4(X_8)+2.2g_5(X_9)+\epsilon,$$ 
	where $g_1(X)=\cos(2X),\ g_2(X)=X^2,\ g_3(X)=X,\ g_4(X)=(X-1)^2$, and $g_5(X)=\sin(X-3).$ Other settings remain to be the same as in Scenario 1. The simulation results in Tables \ref{tab:Table1} demonstrate the superiority of our proposed method in detecting non-linear and non-monotonic relationships compared to other methods, particularly in complex transformation models. Additionally, the results show that our method is robust under different error distributions. In contrast, other methods fail to achieve sure screening property and completely miss variables such as $X_1$ and $X_2$, which are actually related to the outcome in a non-monotonic fashion. It is worth noting that although SII can also accommodate nonlinearity relationships, it considers the difference in survival probability instead of RMST, which can lead to increased noise in some cases when the signal is not strong enough. 
 
\begin{table}[ht]
    \caption{Simulation results for linear transformation model in Scenario 1 and nonlinear additive model in Scenario 2 on Median and IQR of the model size, and the coverage probability.}
    \label{tab:Table1}
    \centering
    \scalebox{0.75}{
    \begin{tabular}{ccccccccccc}
			\hline
			       & &    &        & Median & IQR & Pall & Median & IQR & Pall\\
			Scenario & Error & CR & Method & n=100  &     &      & n=200  &     &     \\
			\hline
			Scenario 1 & N(0,1) & 20\% & RMST & 4      & 2   & 93\% & 4      & 0   & 100\%\\
			   &    &      & CRIS & 4      & 1   & 97\% & 4      & 0   & 100\%\\
			    &   &      & CRSIS& 4      & 1   & 97\% & 4      & 0   & 100\%\\
			  &     &      & SII  & 24     & 52  & 46\% &6       & 3   & 100\%\\
			  &     &      & LASSO& -      & -   & 100\%& -      & -   & 100\%\\
			  &     & 40\% & RMST & 6      & 7   & 82\% & 4      & 0   & 100\%\\
			  &     &      & CRIS & 5      & 4   & 86\% & 4      & 0   & 100\%\\
			  &     &      & CRSIS& 6      & 9   & 86\% & 4      & 0   & 100\%\\
			  &     &      & SII  & 23     & 72  & 43\% &5       & 3   & 100\%\\
			  &     &      & LASSO& -      & -   & 100\%& -      & -   & 100\%\\
			& Standard Extreme & 20\% & RMST & 4      & 2   & 97\% & 4      & 0   & 100\%\\
			&&      & CRIS & 4      & 3   & 95\% & 4      & 0   & 100\%\\
			&&      & CRSIS& 4      & 3   & 93\% & 4      & 0   & 100\%\\
			&&      & SII  & 23     & 45  & 48\% &6       & 6   & 100\%\\
			& &     & LASSO& -      & -   & 100\%& -      & -   & 100\%\\
			& & 40\% & RMST & 5      &5    & 84\% & 4      & 0   & 100\%\\
			&   &   & CRIS & 7      & 6   & 81\% & 4      & 0   & 100\%\\
			&    &  & CRSIS& 6      & 6   & 82\% & 4      & 0   & 100\%\\
			&     & & SII  & 39     & 136 & 33\% &7       & 6   & 96\%\\
			&      && LASSO& -      & -   & 100\%& -      & -   & 100\%\\
		&	Logistic& 20\% & RMST & 6      &13   & 73\% & 4      & 0   & 100\%\\
		&	&      & CRIS & 5      &10   & 78\% & 4      & 1   & 99\%\\
		&	&      & CRSIS& 5      &11   & 77\% & 4      & 0   & 100\%\\
		&	&      & SII  & 79     &253  & 26\% &9       &18   &  88\%\\
		&	&      & LASSO& -      & -   & 100\%& -      & -   & 100\%\\
		&	& 40\% & RMST &30      &100  & 46\% & 4      & 0   & 100\%\\
		&	&      & CRIS & 14      &56   &56\% & 4      & 0   & 100\%\\
		&	&      & CRSIS& 13     &33   & 65\% & 4      & 0   & 100\%\\
		&	&      & SII  & 53     &144  & 31\% &8       & 8   & 100\%\\
		&	&      & LASSO& -      & -   & 100\%& -      & -   & 100\%\\
			\hline
   Scenario 2 & N(0,1) & 20\% & RMST & 41     & 82  & 31\% & 6      & 1   & 98\%\\
		&	&      & CRIS & 689    & 852  & 3\% & 289    & 597  & 19\%\\
		&	&      & CRSIS& 1286   & 941  & 0\% & 1011   & 1187 & 5\%\\
		&	&      & SII  & 192    & 266 & 8\% &22       & 38  & 66\%\\
		&	&      & LASSO& -      & -   & 0\%& -      & -   & 0\%\\
		&	& 40\% & RMST & 178    & 273 & 8\% & 8       & 7   & 94\%\\
		&	&      & CRIS & 1206   & 1105& 0\% & 530     & 749 & 7\%\\
		&	&      & CRSIS& 1075   & 777 & 0\%  & 1008   & 1053& 3\%\\
		&	&      & SII  & 632    & 539 & 0\% &155       & 277   & 15\%\\
		&	&      & LASSO& -      & -   & 0\%& -      & -   & 0\%\\
		&	Standard Extreme & 20\% & RMST & 55     & 137 & 86\% & 5  & 2  & 100\%\\
		&	&      & CRIS &598      & 636& 7 \% & 427    & 503 & 17\%\\
		&	&      & CRSIS& 1195   & 722 & 0\% & 953   & 884  & 3\%\\
		&	&      & SII  & 244     &392  &  0\% &15      & 21  & 79\%\\
		&	&      & LASSO& -      & -   & 0\%& -      & -   & 3\%\\
		&	& 40\% & RMST & 83     &179  & 14\% & 8      & 5   & 96\%\\
		&	&      & CRIS & 752    & 930 & 2\% & 837    & 901  &   9\%\\
		&	&      & CRSIS& 1407   & 705 & 0\% & 1175   & 954   & 4\%\\
		&	&      & SII  & 282    & 323 &  5\% &73     & 195  & 25\%\\
		&	&      & LASSO& -      & -   &  0\%& -      & -   &  0\%\\
		&	Logistic& 20\% & RMST & 46     &112 & 28\% & 7     & 5  & 95\%\\
		&	&      & CRIS & 752    &486  & 2\% & 331    & 709 & 16\%\\
		&	&      & CRSIS& 1240   &785  & 0\% & 1123     &846 & 2\%\\
		&	&      & SII  & 242   &332   & 0\% &25       &54   &  73\%\\
		&	&      & LASSO& -      & -   & 0\%& -      & -   & 1\%\\
		&	& 40\% & RMST &201     &346  & 3\% & 10     &14   & 88\%\\
		&	&      & CRIS & 837    &928   &0\% &958    & 1202  & 1\%\\
		&	&      & CRSIS& 1196   &874  & 0\% &1092   & 989  & 0\%\\
		&	&      & SII  & 254    &369  & 0 \% &112    & 157   & 21\%\\
		&	&      & LASSO& -      & -   & 0\%& -      & -   & 0\%\\
   \hline
		\end{tabular}}
	\end{table}
		
	\textbf{Scenario 3.} In this scenario, we investigate the robustness of the proposed measure by examining a contaminated predictor similar to the Example 3.2 in \cite{hao2019}. All model settings are the same as in Scenario 2, except that $\boldsymbol{X}\sim0.9\boldsymbol{Z}_1+0.1\boldsymbol{Z}_2$, where $\boldsymbol{Z}_1\sim N(0,\Sigma), \Sigma_{ij}=0.5^{|i-j|}$ and $\boldsymbol{Z}_2$ is the $p-$dimensional contamination that follows a $t$ distribution with degrees of freedom 2, independent of $Z_1$. We consider a censoring rate of $20\%$ and sample sizes of $n=200,400$.  The results, summarized in Table~\ref{tab:tab3}, demonstrate that the proposed method exhibits favorable and stable performance over other methods in all cases, even when there are outliers in the covariates.
	
	\textbf{Scenario 4.} To study the existence of local dependency, we consider a change point model here with
	$H(T)=h_1(X_1)+h_2(X_2)+h_3(X_8)+h_4(X_9)+\epsilon,$ where
	\begin{flalign*}
	&&& h_1(x)=\left\{
    \begin{aligned}
		&0.6x^2,\ x\leq -0.8\ or\ x>0.8; \\
		&0,\ \mbox{otherwise};
	\end{aligned}
	\right. 
	&&& h_2(x)=\left\{
	\begin{aligned}
		&1.2x,\ x<-1; \\
		&0,\ \mbox{otherwise};
	\end{aligned}
	\right.
	&&&\ 
	\end{flalign*}
	
	\begin{flalign*}
	&&& h_3(x)=\left\{
	\begin{aligned}
		&0.2,\ x\leq-1\ or\ x\geq0; \\
		&1.8,\ -1<x<0;
	\end{aligned}
	\right.
	&&& h_4(x)=\left\{
	\begin{aligned}
		&1.8,\ -1\leq x\leq 0.5; \\
		&0,\ \mbox{otherwise.}\\
	\end{aligned}
	\right.
	&&&\ 
	\end{flalign*}

	Due to the complexity and local dependencies in the model, we only compare the proposed RMST screening with SII method. The simulation results are summarized in Table~\ref{tab:tab2}. The results demonstrate that the sure screening property of the proposed method is still valid. Moreover, the table shows that the proposed method has good performance in terms of $P_{all}$ and the minimum size required to select all active variables, even when the covariates are only locally related to the survival outcome. Overall, the proposed method outperforms the SII method in this setting.
	
	\textbf{Scenario 5.} In this example, we evaluate the performance of our proposed iterative screening procedure in the presence of high collinearity among the covariates. In the initialization procedure in iterative screening, we adopt RMST screening, CRSIS, CRIS and SII for comparisons. To ensure comparability, we set the number of selected variables at $n/2=50$ and $p=1000$. We consider the same linear transformation model as in Scenario 1 except that $\beta=(2,1.8,\boldsymbol{0}_6,-1.6,-2)'$ and $\boldsymbol{X}=0.9\boldsymbol{Z}_1+0.1\boldsymbol{Z}_2$, where $\boldsymbol{Z}_1\sim N(0,\Sigma),\ \Sigma_{ij}=\rho \ (i\neq j)$ and $\boldsymbol{Z}_2$ is the contamination that follows a multivariate $t$ distribution with degrees of freedom 2 and is independent of $Z_1$, and covariance matrix $\tilde{\Sigma}$, where $\tilde{\Sigma}_{ij}=0.5^{|i-j|}.$ We summarize the results for $\rho=0.5$ and $\rho=0.9$ with normal error term in Table \ref{Tab:Table3}. From the table, we observe that when there is heavy multicollinearity among the covariates, existing methods perform poorly even in the linear scenario. However, the proposed iterative method can improve the screening performance a lot. The iterative screening procedure with RMST initialization is the best among all.
	
	Overall, the above five scenarios demonstrate that the proposed RMST based screening and its iterative version are more sensitive in capturing various global or local dependencies between covariates and the survival outcome, compared to other model-free screening procedures or model-based parametric methods.
	
\textbf{Additional Scenario.} Additional simulation results for interval-censored data are presented in Table S2 in Web Appendix C. In this scenario, inspection times were generated with an interval of 1 unit time, the dimension is $p=1000$, and other settings are the same as in the above Scenarios 1 to 4. We compare our method with \cite{hu2020}, and the results indicate that the proposed RMST-based screening has a better $P_{all}$ of true important variables, especially under logistic error term. This demonstrates the potential and robustness of using RMST-based screening for interval-censored data.

\begin{table}[ht]
	\caption{Simulation results for additive model with contamination in covariates in Scenario 3 on Median and IQR of the model size, and the coverage probability.}
  \label{tab:tab3}
        \centering
        \small
		\begin{tabular}{ccccccccc}
			\hline
			&    &        & Median & IQR & Pall & Median & IQR & Pall\\
			Error & CR & Method & n=200  &     &      & n=400  &     &     \\
			\hline
			N(0,1) & 20\% & RMST & 18     & 89  & 64\% &11      & 11  & 100\%\\
			&      & CRIS & 750    & 904 & 0\% & 353     & 579 & 11\%\\
			&      & CRSIS& 1541   & 731 & 0\% & 1241    & 428 & 9\%\\
			&      & SII  & 67     & 50  & 44\% &36       &11   &86\%\\
			&      & LASSO& -      & -   &  0\%& -      & -   & 0\%\\
			\hline
			Standard Extreme & 20\% & RMST &17    & 27  & 75\% &6      & 1  & 100\%\\
			&      & CRIS & 695    & 854 & 5\% & 648     & 699 & 9\%\\
			&      & CRSIS& 1395   & 656 & 0\% & 1169    & 434& 10\%\\
			&      & SII  & 56     & 61  & 46\% &29       &10   &89\%\\
			&      & LASSO& -      & -   &  0\%& -      & -   & 0\%\\
			\hline
			Logistic & 20\% & RMST &21    &86  & 58\% &10    & 11  & 100\%\\
			&      & CRIS & 704    & 740 & 5\% & 433     & 674 & 0\%\\
			&      & CRSIS& 1421   & 772 & 0\% & 653    &709& 2\%\\
			&      & SII  & 71     & 49  & 39\% &43       &13   &71\%\\
			&      & LASSO& -      & -   &  0\%& -      & -   & 0\%\\
			\hline
		\end{tabular}
	\end{table}

\begin{table}[ht]
	\caption{Simulation results for change point model in Scenario 4 on Median and IQR of the model size, and the coverage probability. }
 \label{tab:tab2}
        \centering
        \small
	\begin{tabular}{cccccc}
		\hline
		&    &        &  Median & IQR & Pall\\
		Error & CR & Method   & n=200  &     &     \\
		\hline
		N(0,1) & 20\% & RMST & 5      & 1   & 100\%\\
		&      & SII   &39       & 36  & 68\%\\
		& 40\% & RMST &  4       & 3   & 100\%\\
		&      & SII  & 48    & 54   & 65\%\\
		\hline
		Standard Extreme & 20\% & RMST & 6 & 1 & 95\%\\
		&      & SII  &75      & 175 & 47\%\\
		& 40\% & RMST &8      &8   & 94\%\\
		&      & SII  &85    & 121  & 42\%\\
		\hline
		Logistic& 20\% & RMST & 10 &14 &95\%\\
		&      & SII  & 132     &320  & 34\%\\
		& 40\% & RMST & 12     &14   & 90\%\\
		&      & SII  & 147    & 128  & 20\%\\
		\hline
	\end{tabular}
\end{table}

\begin{table}[ht]
	\caption{Simulation results for model with high collinearity in Scenario 5 on Median and IQR of the model size, and the coverage probability. 'IrRMST', 'IrCRIS', 'IrCRSIS', and 'IrSII' indicates the iterative screening methods based on different marginal screening methods in the initialization.}
  \label{Tab:Table3} 
		\begin{center}
		\begin{tabular}{cccccc}
			\hline
			Error&  $\rho$   &Method   & Median & IQR & Pall \\
			\hline
			N(0,1) & 0.5 & RMST & 67     & 256  & 24\% \\
			&      & IrRMST& - & -&   94\%\\
			&      & CRIS & 756     & 87 & 0\% \\
                 &      & IrCRIS& - & -&   74\%\\
			&      & CRSIS& 341   & 512 & 13\% \\
                 &      & IrCRSIS& - & -&   90\%\\
			&      & SII  & 617 & 240  & 28\% \\
                 &      & IrSII& - & -&   85\%\\
			&      & LASSO& -      & -   &  87\%\\
			\hline
			 & 0.9 & RMST & 621 & 302  & 4\% \\
			&      & IrRMST& - & -&   84\%\\
			&      & CRIS & 822 & 212 & 0\% \\
                 &      & IrCRIS& - & -&   72\%\\
			&      & CRSIS& 579   & 503 & 0\% \\
                 &      & IrCRSIS& - & -&  80\%\\
			&      & SII  & 750    & 193  & 2\% \\
                  &      & IrSII & - & -&   64\%\\
			&      & LASSO& -      & -   &  38\%\\
			\hline
		\end{tabular}
		\end{center}
	\end{table}

	\section{A Breast Cancer Example}
	Recent advances in personalized treatment call for accurate prognostic system to offer therapeutic options for breast cancer. However, conventional studies aimed at identifying gene signatures for predicting risk of breast cancer have mostly relied on simple and multivariable Cox regression models, which only consider biomarkers with linear effects on survival response. This results in an unsatisfactory prognostic system \cite{bao2019,du2019,kuang2020}. To fill the gap, we applied the proposed screening method to the breast cancer dataset from \cite{goldman2019} to identify features with potential nonlinear association with the response and construct a risk score for the prognosis system of breast cancer with the aid of the generalized additive Cox model.
	
\begin{figure}[ht] 
		\centering 		\includegraphics[height=1\textwidth,width=0.55\textwidth,angle=90]{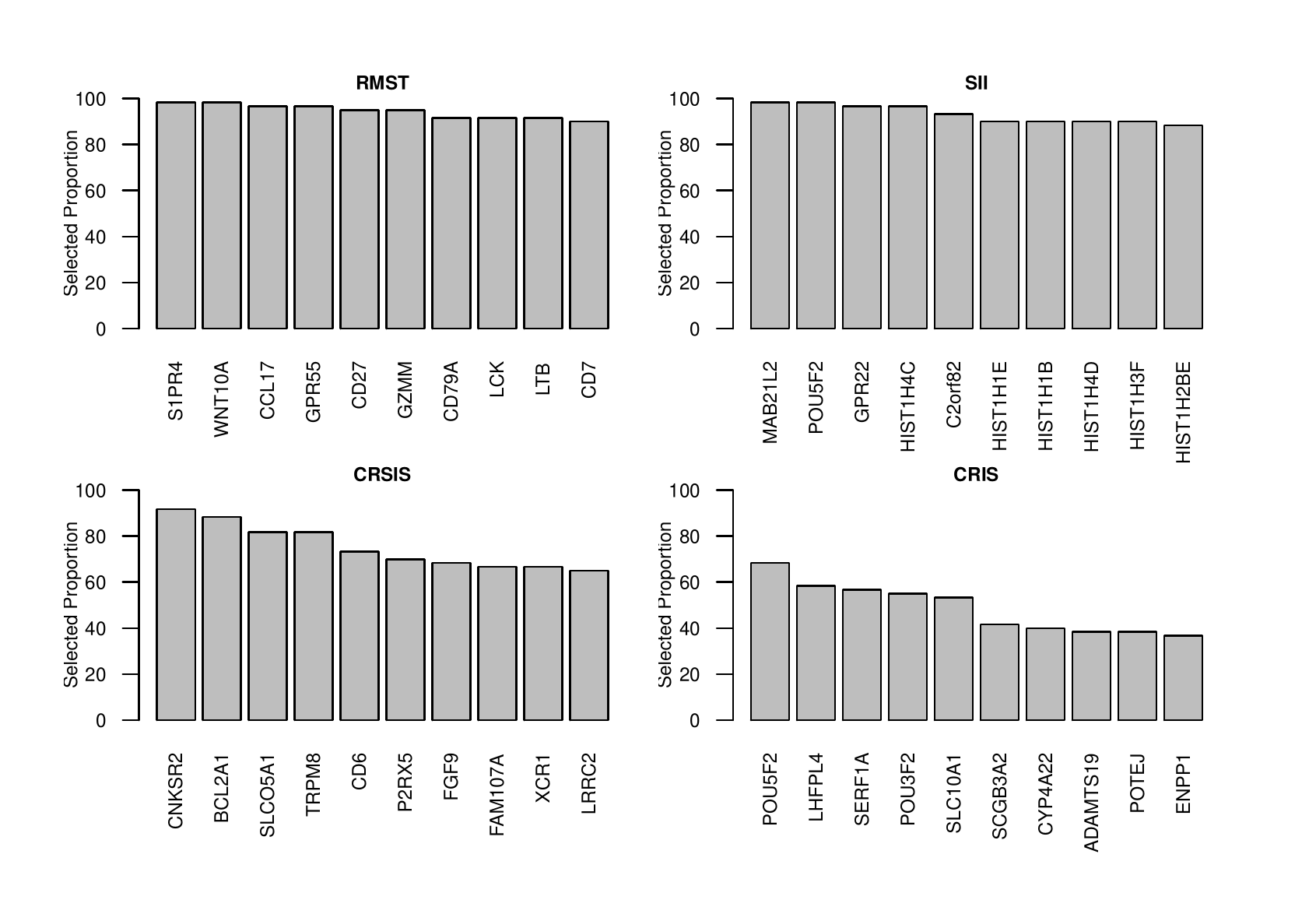} 

\includegraphics[height=1\textwidth,width=0.55\textwidth,angle=90]{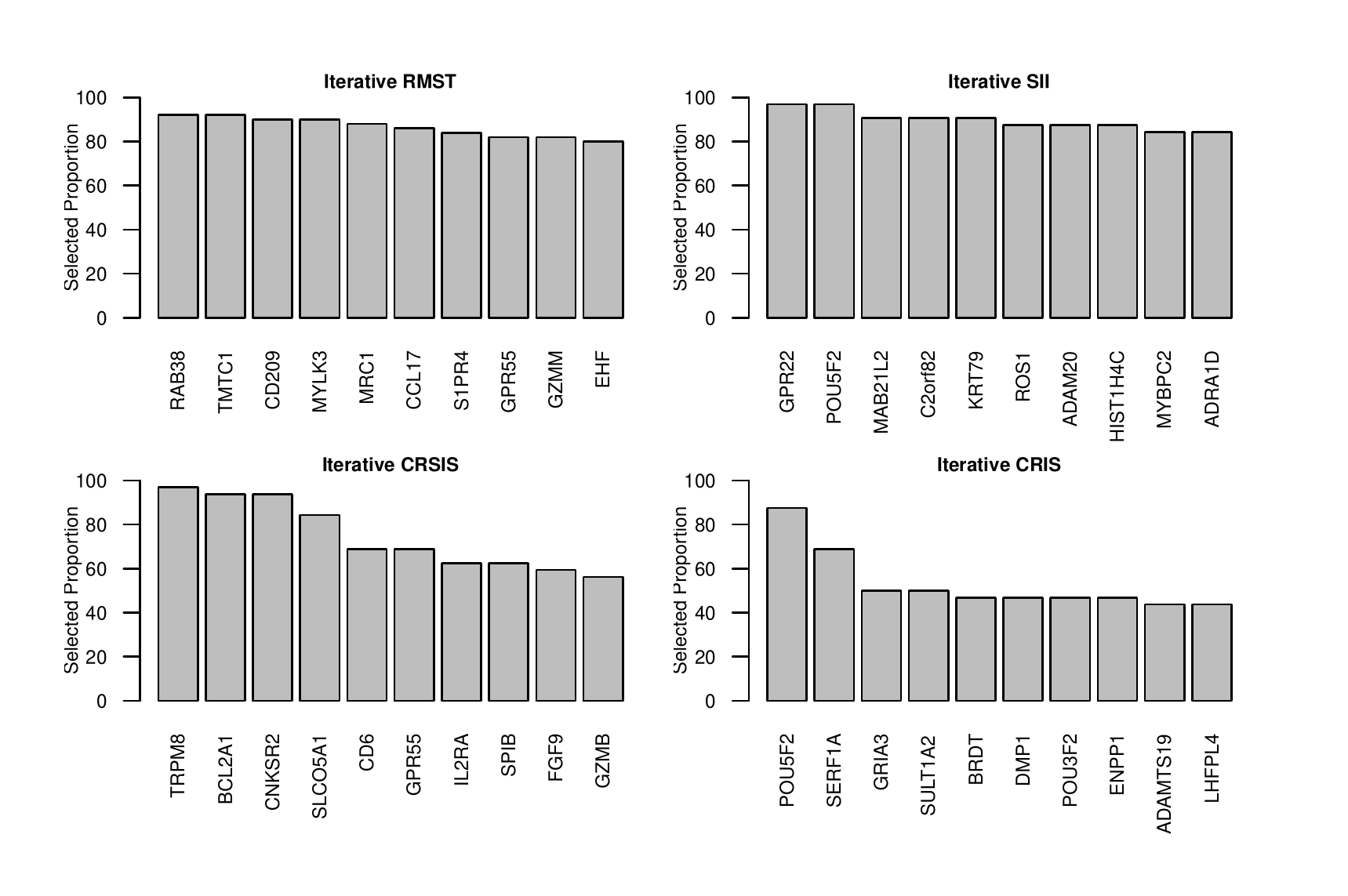} 
		
		\caption{Top 10 genes in the breast cancer data set selected by different screening methods.} 
        \label{fig:1}
	\end{figure}

	We downloaded level 3 gene expression data of breast cancer (BRCA) from The Cancer Genome Atlas (TCGA) project on the UCSC Xena platform \cite{goldman2019}. The dataset consisted of 58387 genes, of which we considered only 18781 protein-coding genes in our analysis. Patient samples without paracarcinoma tissue were excluded, leaving us with 111 patients. The variable of interest is the overall survival of the patients and the survival time is right-censored. Following \cite{bijlsma2006}, we eliminated genes with more than 20$\%$ missing samples, and we retained highly informative genes with the $20\%$ largest coefficient of determination. This resulted in a candidate gene set of $p_n=$3189 genes and $n=$111 observations. The data are randomly divided into a training set with $n_{train}=90$ samples and a testing set with $n_{test}=21$ samples. We then applied the proposed RMST-based marginal and iterative screening to the candidate gene set and compared it with the well-established SII-based screening, CRSIS, CRIS, all with the same size of $\gamma_n=2[n/\log n]$, as well as the corresponding iterative screening. To ensure the stability and robustness of the gene set, we replicated the experiment 100 times and analyzed the results using an ensemble approach. We recorded the proportion of times among 100 replications that each gene was selected and denote them by $\hat{\eta}=(\hat{\eta}_1,\cdots,\hat{\eta}_{p_n})'$ where $\hat{\eta}_j=\frac{1}{100}\sum_{k=1}^{100}I(j\in\mathcal{P}_k),\ j=1,\cdots,p_n$. Genes are sorted in descending order based on their selected proportions, and the top 10 genes were summarized in Figure \ref{fig:1}. 
	

	The results presented in Figure \ref{fig:1} show that RMST and SII based marginal screening methods and iterative RMST screening are more robust than CRIS and CRSIS, as the top 10 selected genes are highly consistent across the 100 data replications. Upon comparing RMST and Iterative RMST as well as SII and Iterative SII, we can see that RMST-based screening is more effective as four genes (CCL17, GPR55, S1PR4, and GZMM) appear in both Iterative RMST and RMST, indicating the reliability of these potential gene signatures. 
   
	Next, we evaluated the predictive power of the selected genes by C-index \cite{Pencina2004} and the risk score. The overall C-index measures the concordance between the predicted risk score and the actual survival time, where a higher C-index indicates better performance, and it is defined as $P(\hat{\lambda}_i>\hat{\lambda}_j|Y_i>Y_j)$, where $\hat{\lambda}_i$ and $\hat{\lambda}_j$ are the estimated risk score of subjects $i$ and $j$. Specifically, we fit the generalized Cox additive model with the top $[n/\log n/3]=7$ selected genes from each method in Figure~\ref{fig:1} on the training set, as the degrees of freedom of the spline in the model is 3, and then validate the estimation on testing set by computing the risk score and C-index. Table \ref{Tab:Table5} summarizes the average overall C-index and its standard deviation among 100 random split of the training and testing set. In all, iterative screening can help to improve the predictive power. Among the marginal methods, RMST screening outperforms its counterparts with higher C-index, and among the iterative methods, Iterative RMST screening is also the best. These findings suggest that the proposed RMST-based screening method and the iterative screening procedure can effectively identify potential gene signatures and build a robust and more reliable prognostic system for breast cancer.
	
\begin{figure}[ht]
    \centering \includegraphics[scale=0.4]{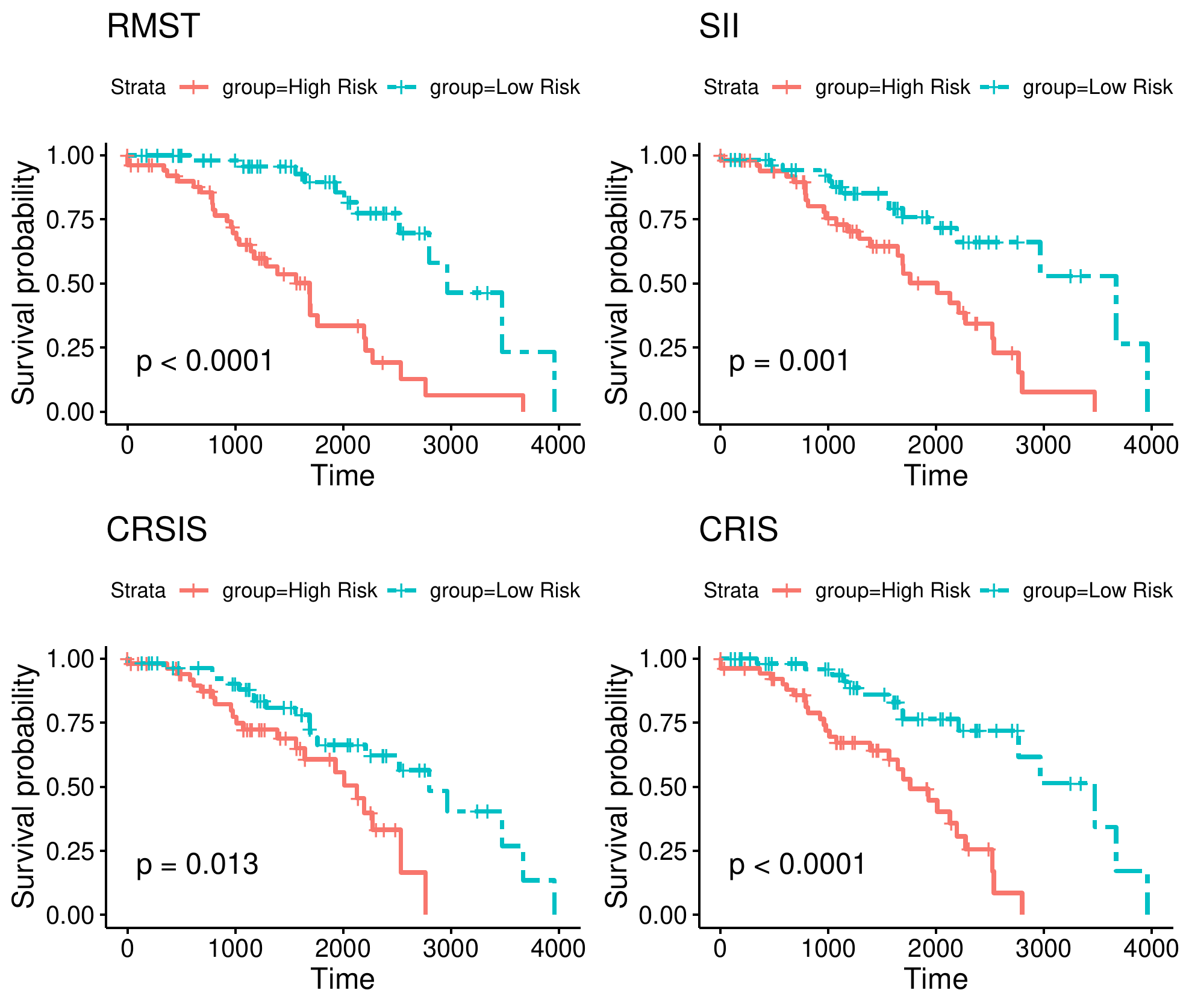} 
    \includegraphics[scale=0.4]{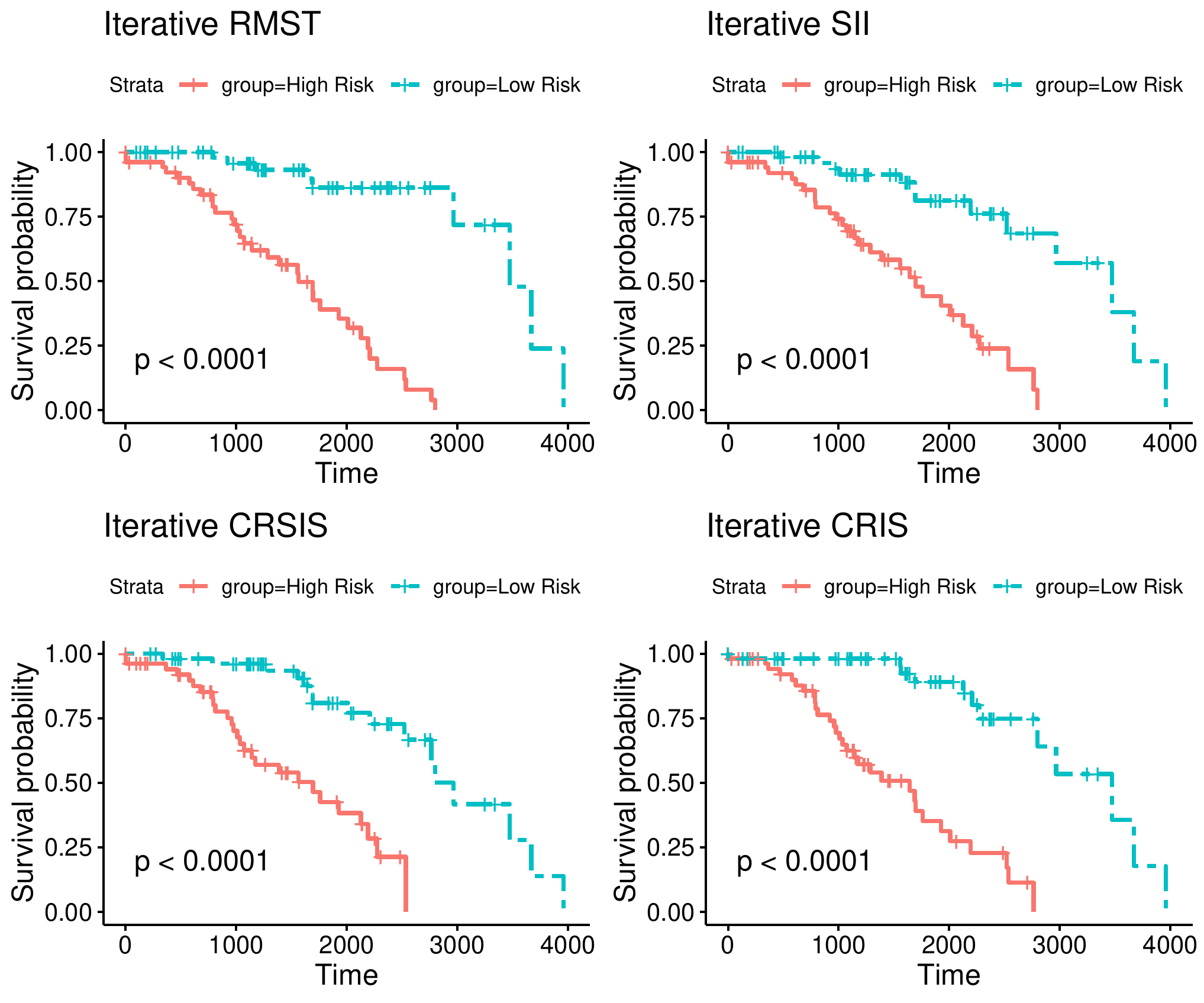} 
    \caption{KM estimator of the survival curves under predicted high risk group and low risk group classified by predicted median risk score.} 
  \label{Fig.main3} 
	\end{figure}

    Figure \ref{Fig.main3} shows the prognostic performance of the GAM model using the top 7 variables selected from different methods in Figure~\ref{fig:1}. The samples were classified into higher and lower risk groups based on the predicted median risk score. The results indicate that the genes selected by iterative screening with RMST initialization is the most accurate to make a prognosis of the risk of breast cancer, as it has the smallest log-rank test p-value of $10^{-10}$.
	    
	After examining the fitted splines of the GAM model among the top 20 related genes selected by the proposed method, we have identified S1PR4 and HSD11B1 as potential genes with significant nonlinear and nonmonotonic effects on the restricted mean survival time. As shown in Figure S2 in the supplementary materials, a moderate expression level of S1PR4 is associated with a higher risk of death, while HSD11B1 shows the opposite effect. It is worth noting that HSD11B1 has been reported to be related to better disease-free survival in breast cancer \cite{vishnubalaji2020}, while S1PR4, a member of the S1PRs family, is known to be associated with suppressing disease progress in breast cancer patients \cite{lei2018}. Further studies are needed to confirm whether there is a nonlinear association between these gene signatures and the prognosis of breast cancer patients. Interestingly, BRCA1 and BRCA2, the most common cause of hereditary breast cancer, were not selected by any of the screening methods. They may not have any effect on survival but just the cause of the breast cancer. The finding is consistent with a recent cohort study that reported no significant difference in overall survival between BRCA-positive and BRCA-negative breast cancer patients \cite{copson2018}.

\begin{figure}[ht]
     \includegraphics[scale=0.5]{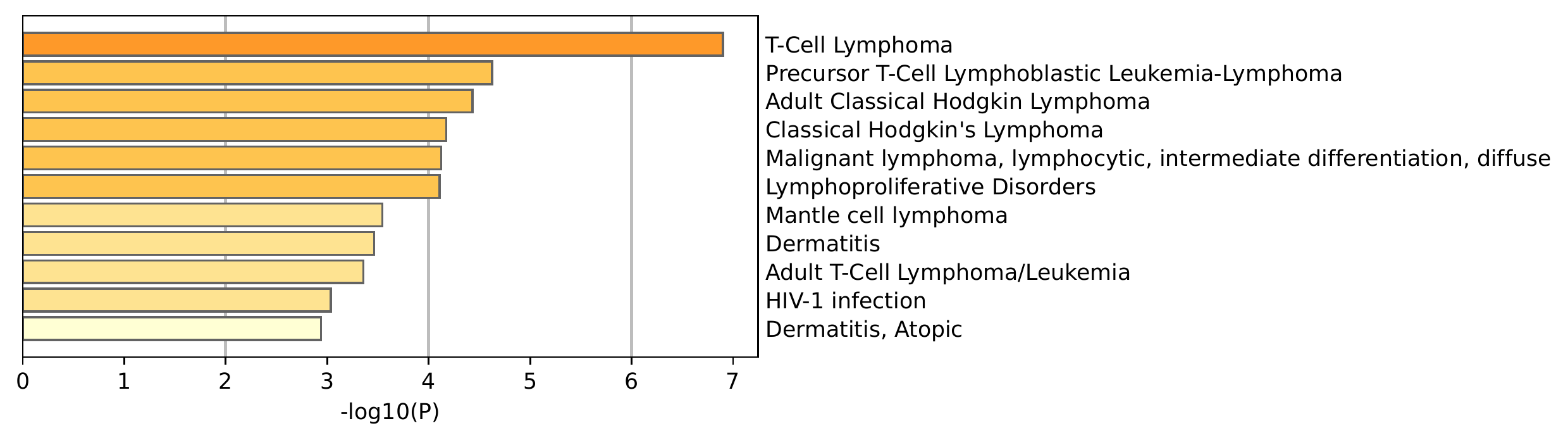} 
    \caption{Summary of enrichment analysis in DisGeNET using 10 selected genes from RMST screening.} 
  \label{Fig.dis1}
	\end{figure}

\begin{figure}[ht]
   \includegraphics[scale=0.5]{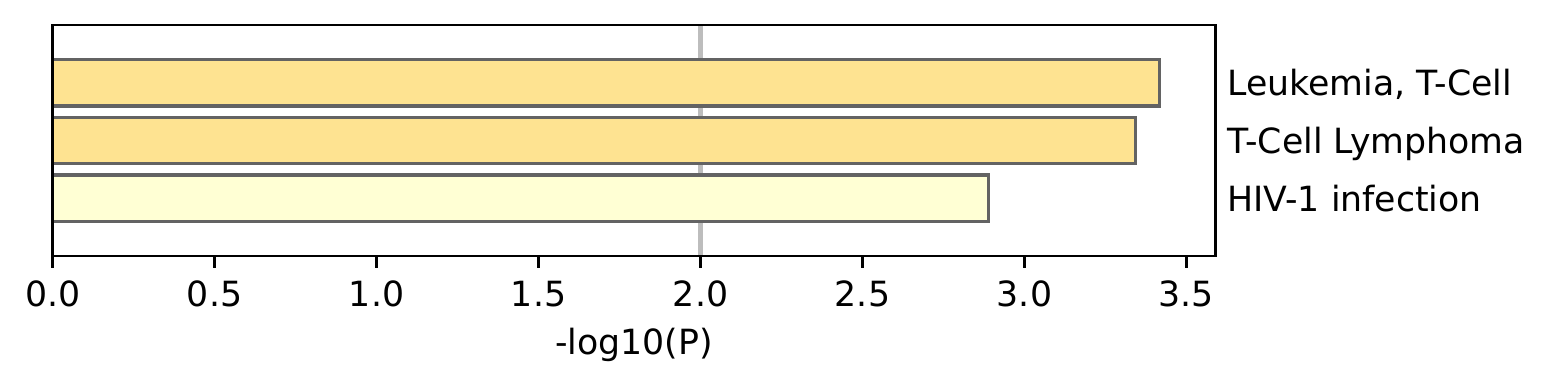}
    \caption{Summary of enrichment analysis in DisGeNET using 10 selected genes from Iterative RMST.} 
  \label{Fig.dis3}
	\end{figure}

In the final step of our analysis, we performed enrichment analysis on the selected gene sets of RMST screening and Iterative RMST screening using DisGeNet on the Metascape platform \cite{zhou2019} to validate their potential relevance. The results are presented in Figures~\ref{Fig.dis1} and \ref{Fig.dis3}. Specifically, the proposed method identified genes that were associated not only with breast cancer, but also with lymphoma-related diseases, leukemia, as well as dermatitis and allergic contact. These findings are consistent with existing research indicating a relationship between breast cancer and lymphoma \cite{Engkildee2011,Bakkach2018,wiernik2000}. Our study provides further evidence for potential common prognostic genes in these diseases, laying a foundation for future research in cancer prevention and treatment.

		\begin{table}[ht]
		\caption{C-index and standard deviation in testing set. 'Ir-' represents iterative screening with corresponding initialization screening. }
  \label{Tab:Table5}
  \centering
			\scalebox{0.9}{\begin{tabular}{cccccccccc}
				\hline
				Method & RMST & SII & CRSIS & CRIS  & IrRMST & IrSII & IrCRSIS & IrCRIS\\
				\hline
				Avg C & 0.644 & 0.560 & 0.553 & 0.619 & 0.749 & 0.610 & 0.689 & 0.659\\
				SD & 0.131 & 0.121 & 0.131 & 0.120 & 0.112 & 0.129 & 0.105 & 0.133\\
				\hline
			\end{tabular}}
	\end{table}

\section{Discussion}
\label{s:discuss}
	This paper presents a novel approach to feature screening for survival data with several contributions to the field. Firstly, we propose a novel approach for evaluating survival-related effects using RMST, which is a more interpretable measure for patients and practitioners in clinical studies. Secondly, the proposed method can detect both linear and nonlinear features with complex non-monotonic patterns based on RMST, outperforming previous methodologies with a theoretically justified sure screening property. We demonstrate the practical value of our method in biomedical research by applying it to a breast cancer example and identifying potential nonlinear gene effects on mean survival time. Simulation studies show that the proposed method performs well on different predictor effect patterns, including quadratic, piece-wise effects, as well as other nonlinear and non-monotonic relationships. Thirdly, we extend our method to an iterative screening procedure to handle high multicollinearity and outliers in predictors, filling a gap in high-dimensional survival data, where less attention has been given to joint analysis. Finally, we illustrate the potential applicability of our screening method in interval-censored data through simulation studies with promising results. In summary, our proposed method has broad implications in biomedical research and can facilitate the identification of potential prognostic features in complex datasets.

    Although the empirical simulation studies for interval-censored data works well, we acknowledge that the theoretical properties of the proposed method have not yet been fully developed due to the complicated asymptotic properties of RMST under interval-censored data. In addition, the proposed method only considered the independent censoring mechanism, and more general censoring mechanisms can be explored in future work. Furthermore, the current screening procedure may not be flexible enough to handle complex dependencies such as network structures. Recently, \cite{tian2021} proposed a joint feature screening framework using random subspace ensemble, but their method is only applicable to uncensored data situations. Inspired by their work, more sophisticated methods can be explored in future studies to address these limitations and improve the applicability of the proposed method in more complex scenarios.

\section*{Supplementary material}
\label{SM}
Supplementary material available online includes the proof of Theorems 1 and 2 and Corollary 1, a motivating toy example, additional simulation results and additional figures in the breast cancer example.

\end{document}


\maketitle
\section{Web Appendix A}
	\textbf{Proof of Theorem 1}
	Since the function $S$ is a functional of integrated hazard $A$, i.e., $S=S(A)=\prod(1-dA)$, and $(dS(A)h)(s)=-S(s)h(s)$, by applying functional delta method, we have that $\hat{S}-S$ is asymptotically equivalent to $-S(\hat{A}-A)$, where $\hat{A}$ is the Nelson-Aalen estimator. Hence, we can find the asymptotic equivalent form of $\hat{S}_j(t;x)-\hat{S}(t)-(S_j(t;x)-S(t))$ as 
	$$\sum_{i=1}^n\big[f_t(Y_i,X_{ji},\Delta_i;t,x)+g_t(Y_i,\Delta_i;t)\big],$$ where
	\begin{equation*}
		\begin{split}
		f_t(Y_i,X_{ji},\Delta_i;t,x)=&-S_j(t;x)\int_0^t \frac{J_i(s|X_j>x)}{Y(s|X_j>x)}dM(s)\\
	 	&=-S_j(t;x)\int_0^t\frac{J_i(s|X_j>x)}{Y(s|X_j>x)}dM(s)\\
		&=-S_j(t;x)I(X_{ji}>x)\big[\int_0^t\frac{J_i(s)}{Y(s|X_j>x)}dN(s)-\int_0^tJ_i(s)\alpha_j(s)ds\big],
		\end{split}
	\end{equation*}
	and
		\begin{equation*}
		\begin{split}
			g_t(Y_i,\Delta_i;t)=-S(t)\big[\int_0^t\frac{J_i(s)}{Y(s)}dN_i(s)-\int_0^tJ_i(s)\alpha(s)ds\big],
		\end{split}
	\end{equation*}
	where $Y(s)=I(Y_i>s)$, $J_i(s)=I(Y_i(s)>0)$, $N_i(s)=I(Y_i\leq s,\Delta_i=1)$, and $M(s)$ is a counting process martingale.
	
	The class of indicator functions $\{f_t(Y_i,X_{ji},\Delta_i;t,x), (t,x)\in \mathcal{T} \times \mathcal{X}_j\}$, and $\{g_t(Y_i,\Delta_i;t), (t,x)\in \mathcal{T} \times \mathcal{X}_j\}$ are both Donskers. Also, denote $h_t(Y_i,X_{ji},\Delta_i;t,x)=f_t(Y_i,X_{ji},\Delta_i;t,x)+g_t(Y_i,\Delta_i;t)$, then  $\{h_t(Y_i,X_{ji},\Delta_i;t,x), (t,x)\in \mathcal{T} \times \mathcal{X}_j\}$ is Donsker. Therefore, according to Donsker Theorem (Functional Central Limit Theorem), we have 
	\begin{equation}
	\sqrt{n}\big[\hat{S}_j(t;x)-\hat{S}(t)-(S_j(t;x)-S(t))\big]\stackrel{\mathrm{w}^*}{\to}\mathbbm{G}_j(t,x),
	\end{equation}
	$\mathbbm{G}_j(t,x)$ is a Brownian bridge with mean 0 and covariance matrix $Cov(\mathbbm{G}_j(t,x_1),\mathbbm{G}_j(s,x_2))=E\big[\{h_t(Y_i,X_{ji},\Delta_i;t,x_1)\{h_t(Y_i,X_{ji},\Delta_i;s,x_2)\big]$.
	
	If $d_{1j}=0$, i.e., predictor $X_j$ has nothing to do with the survival function, then the Lebesgue measure of $\{(t,x):|S(t;x)-S(t)|\neq0,(t,x)\in \mathcal{T}\times \mathcal{X}_j\}$ is 0. Then equation (1) becomes 
	\begin{equation}
		\sqrt{n}\big[\hat{S}_j(t;x)-\hat{S}(t))\big]\stackrel{\mathrm{w}^*}{\to}\mathbbm{G}_j(t,x).
	\end{equation}
	By continuous mapping theorem, let $g(\cdot)=\int_{\mathcal{X}_j}|\int_{\mathcal{T}} \cdot dt|dF_j(x)$, we conclude that 
	$$\sqrt{n}\hat{d}_{1j}=\sqrt{n}\int_{\mathcal{X}_j}\big|\hat{RMST}_j(x)-\hat{RMST}\big|dF_j(x)\stackrel{\mathrm{w}^*}{\to}\int_{\mathcal{X}_j}\big|\int_{\mathcal{T}} \mathbbm{G}_j(t,x) dt\big|dF_j(x).$$

	When $d_{1j}\neq 0$ and the Lebesgue measure of $\Omega(\kappa_j)=\{(t,x):|\int_\mathcal{T}S_j(t;x)-S(t)dt|<\kappa_j,(t,x)\in\mathcal{T}\times\mathcal{X}_j\}$ is 0, let $\phi(\cdot)=\int_{\mathcal{X}_j}|\int_{\mathcal{T}}\cdot dt|dF_j(x)$ being a map$\ \mathbbm{D}_\phi\subset\mathbbm{D}\to\mathbbm{E}$. Then, $\phi(\cdot)$ is Hadamard-differentiable at $\theta = S_j(\cdot,\cdot)-S(\cdot)$. In fact, there is a continuous linear map $\phi'_\theta:\mathbbm{D}\to\mathbbm{E}$ such that 
	\begin{equation*}
		\frac{\phi(\theta+t_nh_n)-\phi(\theta)}{t_n}\ {\to}\ \phi'_\theta\ =\ \int_{\mathcal{X}_j}sgn\bigg(\int_{\mathcal{T}}\theta dt\bigg)\int_{\mathcal{T}}hdtdF_j(x),  
	\end{equation*}
	for all converging sequence $t_n\to0$ and $h_n\to h$ such that $\theta+t_nh_n\in \mathbbm{D}_\phi$ for every $n$. Finally, applying the functional Delta method, we have
	\begin{equation}
		\begin{split}
		\sqrt{n}(\hat{d}_{1j}-d_{1j})=\sqrt{n}\int_{\mathcal{X}_j}\big|\hat{RMST}_j(x)-\hat{RMST}-\big(RMST_j(x)-RMST\big)|dF_j(x)\\
		\stackrel{\mathrm{w}^*}{\to} \int_{\mathcal{X}_j}sgn\bigg(RMST_j(x)-RMST\bigg)\int_{\mathcal{T}}\mathbbm{G}_j(t,x)dtdF_j(x).
		\end{split}
	\end{equation}
	
	Note that the right-hand side is actually $E_{X_j}\big[sgn\big(RMST_j(x)-RMST\big)\int_{\mathcal{T}}\mathbbm{G}_j(t,x)dt\big]$ in (3). However, when the null hypothesis is true, the limiting distribution is $E_{X_j}\big[|\int_{\mathcal{T}}\mathbbm{G}_j(t,x)dt|\big]$. Notice that $\int_{\mathcal{T}}\mathbbm{G}_j(t,x)$ is also a Brownian motion with regard to $\sup\{t:t\in\mathcal{T}\}$ while $|\int_{\mathcal{T}}\mathbbm{G}_j(t,x)|$ is a folded normal distribution fixing $\mathcal{T}$. This discrepancy in the 2 limiting distributions constitutes the difference between null and alternative hypothese. 
	
	Similarly, we also have 
	\begin{equation}
		\begin{split}
	\sqrt{n}(\hat{d}_{2j}-d_{2j})=\sqrt{n}\int_{\mathcal{X}_j}\big|\hat{RMST}_{j-}(x)-\hat{RMST}-\big(RMST_{j-}(x)-RMST\big)|dF_j(x)\\
	\stackrel{\mathrm{w}^*}{\to} \int_{\mathcal{X}_j}sgn\bigg(RMST_{j-}(x)-RMST\bigg)\int_{\mathcal{T}}\mathbbm{H}_j(t,x)dtdF_j(x),
		\end{split}
	\end{equation}
	where $\mathbbm{H}_j(t,x)$ is also a Brownian bridge with mean 0 but it is correlated with $\mathbbm{G}_j(t,x)$.\\
	\rightline{\QEDclosed}

\begin{Lemma}[\cite{foldes1981}]
	For some universal constants $c_1$ and $c_2$, and $12/(n\gamma^4)$\ $<\epsilon<1$,
	$$Pr\biggl(\sup_{t\in\mathcal{T}}|\hat{S}(t)-S(t)|>\epsilon\biggl)\leq c_1\exp\{-n\gamma^6\epsilon^2/c_2-\log\epsilon\}.$$    
\end{Lemma}

\begin{Lemma}[\cite{li2016}]
	Let $S_j(\cdot)$ and $S_C(\cdot)$ be the marginal survival function of $X_j$ and $C$, respectively. Suppose that given $t\in\mathcal{T},x_0\in\mathcal{X},$ there exists constants $\gamma,\tau$, and $\lambda$, such that $0<\gamma<S(t_0)S_C(t_0),0<\tau\leq S_j(x_0)$ and $0<\lambda\leq Pr(Y>t_0,X_j>x_0)$ for all $1\leq j \leq p_n$. Then, given $24/(n\tau\gamma^4)<\epsilon<1$,
	$$Pr\biggl(\sup_{t\in\mathcal{T},x\in\mathcal{X}_j}\big|\hat{S}_{j}(t;x)-S_j(t;x)\big|>\epsilon\biggl)\leq c_3\exp\{-nc_4\epsilon^2-c_5\log\epsilon\},$$
	where $c_3,c_4,c_5$ are some constants relying on $\tau,\gamma,\lambda$ when $n$ is sufficiently large.
\end{Lemma}	

	\begin{Lemma}
        Under Condition 1, there exists some constants $b_1,b_2,b_3,b_4$, such that for any $\epsilon>0$, we have
		$$Pr\biggl(\max_{1\leq j \leq p_n}|\hat{d}_j-d_j|>\epsilon\bigg)\leq  p_nb_1\exp\{-nb_2\epsilon^2-b_3\log(b_4\epsilon)\}.$$
		Then, under Conditions 1 and 2, for $n>\frac{1}{m^2}$, we have the following ranking consistency property
		$$Pr\biggl(\max_{j\in\bar{\mathcal{M}}_*}\hat{d}_j\geq\min_{j\in\mathcal{M}_*}\hat{d}_j\biggl)\leq p_nb_1\exp\{-nb_2m^2/4-b_3\log(b_4m/2)\}.$$
	\end{Lemma}
	
	The first result in Lemma 1 shows that the estimator $\hat{d}_j$ is consistent and the second result further illustrates the ranking consistency property by showing that all active variables appear earlier than those inactive ones. 
	
\textbf{Proof of Lemma 1:}

For any $\epsilon>0$, we have
\begin{equation}
	\begin{split}
	&\ Pr(\max_{1\leq j\leq p_n}|\hat{d}_j-d_j|>\epsilon)\\
	=&\ Pr\biggl(\max_{1\leq j\leq p_n}\bigg|\frac{1}{n}\sum_{i=1}^n\big|\int_0^{t_{ji}^1}\hat{S}_{j1}(t;x_{ji})dt-\int_0^{t^*}\hat{S}(t)dt\big|-\int_{\mathcal{X}_j}\big|\int_0^{t_{ji}^1}S_{j1}(t;x)dt-\int_0^{t^*}S(t)dt\big|dF_j(x)\\
	&\ \ \ \ \     +\frac{1}{n}\sum_{i=1}^n\big|\int_0^{t_{ji}^2}\hat{S}_{j2}(t;x_{ji})dt-\int_0^{t^*}\hat{S}(t)dt\big|-\int_{\mathcal{X}_j}\big|\int_0^{t_{ji}^2}S_{j2}(t;x)dt-\int_0^{t^*}S(t)dt\big|dF_j(x)\bigg|>\epsilon\biggl)\\
	\leq&\ Pr\biggl(\max_{1\leq j\leq p_n}\bigg|\frac{1}{n}\sum_{i=1}^n\big|\int_0^{t_{ji}^1}\hat{S}_{j1}(t;x_{ji})dt-\int_0^{t^*}\hat{S}(t)dt\big|-\int_{\mathcal{X}_j}\big|\int_0^{t_{ji}^1}S_{j1}(t;x)dt-\int_0^{t^*}S(t)dt\big|dF_j(x)\bigg|\\
	&\ \ \ \ \ \ \ \ \ \ \ >\frac{\epsilon}{2}\biggl)
	+Pr\biggl(\max_{1\leq j\leq p_n}\bigg|\frac{1}{n}\sum_{i=1}^n\big|\int_0^{t_{ji}^2}\hat{S}_{j2}(t;x_{ji})dt-\int_0^{t^*}\hat{S}(t)dt\big|-\\
	&\ \ \ \ \ \ \ \ \ \ \ \ \ \ \  \int_{\mathcal{X}_j}\big|\int_0^{t_{ji}^2}S_{j2}(t;x)dt-\int_0^{t^*}S(t)dt\big|dF_j(x)\bigg|>\frac{\epsilon}{2}\biggl)\\
	\equiv&\  A_1+A_2.
	\end{split}
\end{equation}
Now, we would evaluate $A_1$ first and $A_2$ is its analogy. To the ease of notation, let $\hat{R}_{1i}=\int_{0}^{t_{ji}^1}\hat{S}_j(t;x_{ji})dt$, $\hat{R}=\int_0^{t^*}\hat{S}(t)dt$, $R_{1i}=\int_0^{t_{ji}^1}S_j(t;x_{ji})dt$, and $R=\int_{0}^{t^*}S(t)dt$. Then,
\begin{equation}
	\begin{split}
		&Pr\biggl(\bigg|\frac{1}{n}\sum_{i=1}^n\big|\int_0^{t_{ji}^1}\hat{S}_{j1}(t;x_{ji})dt-\int_0^{t^*}\hat{S}(t)dt\big|-\int_{\mathcal{X}_j}\big|\int_0^{t_{ji}^1}S_{j1}(t;x)dt-\int_0^{t^*}S(t)dt\big|dF_j(x)\bigg|>\frac{\epsilon}{2}\biggl)\\
		=\  &Pr\biggl(\bigg|\frac{1}{n}\sum_{i=1}^n\big|\hat{R}_{1i}-\hat{R}\big|-\int_{\mathcal{X}_j}\big|R_{1i}-R\big|dF_j(x)\bigg|>\frac{\epsilon}{2}\biggl)\\
		=\ &Pr\biggl(\bigg|\frac{1}{n}\sum_{i=1}^n\big|\hat{R}_{1i}-\hat{R}\big|-\frac{1}{n}\sum_{i=1}^n\big|R_{1i}-R\big|+\frac{1}{n}\sum_{i=1}^n\big|R_{1i}-R\big|-E_{\mathcal{X}_j}\big(\big|R_{1i}-R|\big)\bigg|>\frac{\epsilon}{2}\biggl)\\
		\leq\ &Pr\biggl(\bigg|\frac{1}{n}\sum_{i=1}^n\big(\big|\hat{R}_{1i}-\hat{R}|-|R_{1i}-R|\big)\bigg|>\frac{\epsilon}{4}\biggl)+Pr\biggl(\bigg|\frac{1}{n}\sum_{i=1}^n\big|R_{1i}-R\big|-E_{\mathcal{X}_j}\big(\big|R_{1i}-R\big|\big)\bigg|>\frac{\epsilon}{4}\biggl)\\
		\leq\ &Pr\biggl(\frac{1}{n}\sum_{i=1}^n\big|\hat{R}_{1i}-R_{1i}|>\frac{\epsilon}{8}\biggl)+Pr\biggl(|\hat{R}-R|>\frac{\epsilon}{8}\biggl)+Pr\biggl(\bigg|\frac{1}{n}\sum_{i=1}^n\big|R_{1i}-R\big|-E_{\mathcal{X}_j}\big(\big|R_{1i}-R\big|\big)\bigg|>\frac{\epsilon}{4}\biggl)\\
        \equiv\ & B_1+B_2+B_3. 		
	\end{split}
\end{equation}

The first and second inequalities hold both by Triangular inequality. For $B_1$, applying mean value theorem of the integration, we have
\begin{equation}
	\begin{split}
		&B_1\leq Pr\biggl(\frac{1}{n}\sum_{i=1}^nt_{ji}^* \sup_{t\in\mathcal{T}_i}|\hat{S}_j(t;x_{ji})-S_j(t;x_{ji})|>\frac{\epsilon}{8}\biggl)\\
		\leq\ &Pr\biggl(\max_{1\leq i\leq n}\sup_{t\in\mathcal{T}_i}|\hat{S}_j(t;x_{ji})-S_j(t;x_{ji})|>\frac{\epsilon}{8t^*}\biggl)\\
		\leq\ & Pr\biggl(\sup_{x\in\mathcal{X}_j,t\in\mathcal{T}}|\hat{S}_j(t;x)-S_j(t;x)|>\frac{\epsilon}{8t^*}\biggl)\\
		\leq\ & c_3\exp\{-nc_4\epsilon^2/(64t^{*2})-c_5\log\epsilon/(8t^{*})\},
	\end{split}
\end{equation}
where $t^*=\max\{t:t\in\mathcal{T}\}$ and the last inequality is followed by Lemma A.2. Similarly, following from Lemma A.1, for $B_2$, we have that
\begin{equation}
	\begin{split}
		&B_2\leq Pr\biggl(\int_{t\in\mathcal{T}}|\hat{S}(t)-S(t)|dt>\frac{\epsilon}{8}\biggl)\\
		\leq\ &Pr\biggl(\sup_{t\in\mathcal{T}}|\hat{S}(t)-S(t)|>\frac{\epsilon}{8t^*}\biggl)\\
		\leq\ &c_1\exp\{-n\gamma^6\epsilon^2/(64c_2t^{*2})-\log(\epsilon/8t^*)\}.
	\end{split}
\end{equation}

For $B_3$, we apply Hoeffding's inequality and obtain that 
\begin{equation}
	B_3\leq 2\exp\{-n\epsilon^2/(8M^2)\},
\end{equation}
where $M=\max_{1\leq i\leq n}{|R_{1i}-R|}$ is bounded.

Combing equations (7), (8) and (9), we have that
\begin{equation*}
\begin{split}
     A_1\leq & p_n(c_1\exp\{-n\gamma^6\epsilon^2/(64c_2t^{*2})-\log(\epsilon/8t^*)\}+c_3\exp\{-nc_4\epsilon^2/(64t^{*2})-\\
     & c_5\log(\epsilon/8t^{*})\}+2\exp\{-n\epsilon^2/(8M^2)\}),
\end{split}
\end{equation*}

i.e.,
$$A_1\leq p_nc_6\exp\{-nc_7\epsilon^2-c_8\log(c_9\epsilon)\},$$
where $c_7=\max\{8M^2,\frac{64c_2t^{*2}}{\gamma^6},\frac{64t^{*2}}{c_4}\},$ $c_6=\max\{c_1,c_3,2\}$, $c_8=\max\{c_5,1\}$, $c_9=\frac{1}{8t^*}.$

Similarly,  $A_2\leq p_na_6\exp\{-na_7\epsilon^2-a_8\log(a_9\epsilon)\}$ for some constant $a_6,a_7,a_8,a_9$. Hence,
\begin{equation}
	Pr\biggl(\max_{1\leq i \leq n}|\hat{d}_j-d_j|>\epsilon\bigg)\leq  p_nb_1\exp\{-nb_2\epsilon^2-b_3\log(b_4\epsilon)\}\},
\end{equation}
for some constant $b_1,b_2,b_3$ and $b_4$.

Next, we prove the ranking consistency property based on the above result. Suppose constant $\kappa=\min_{j\in\mathcal{M}_*}d_j-\max_{j\in\bar{\mathcal{M}}_*}d_j>0.$ for $n>\frac{1}{m^2}$, we have 
\begin{equation}
	\begin{split}
		&Pr\biggl(\max_{j\in\bar{\mathcal{M}}_*}\hat{d}_j\geq\min_{j\in\mathcal{M}_*}\hat{d}_j\biggl)=Pr\biggl(\max_{j\in\bar{\mathcal{M}}_*}\hat{d}_j-\max_{j\in\bar{\mathcal{M}_*}}d_j\geq\min_{j\in\mathcal{M}_*}\hat{d}_j-\min_{j\in\mathcal{M}_*}d_j+\kappa\biggl)\\
		\leq\ &Pr\biggl(\max_{j\in\bar{\mathcal{M}}_*}|\hat{d}_j-d_j|-\min_{j\in\mathcal{M}_*}|\hat{d}_j-d_j|\geq\kappa\biggl)\\
		\leq\ &Pr\biggl(\max_{j\in\bar{\mathcal{M}}_*}|\hat{d}_j-d_j|+\max_{j\in\mathcal{M}_*}|\hat{d}_j-d_j|\geq\kappa\biggl)\\
		\leq\ &Pr\biggl(\max_{1\leq j\leq p_n}|\hat{d}_j-d_j|\geq\frac{\kappa}{2}\biggl)\\
		\leq\ &p_nb_1\exp\{-\frac{nb_2\kappa^2}{4}-b_3\log(\frac{b_4\kappa}{2})\}\leq p_nb_1\exp\{-nb_2m^2/4-b_3\log(b_4m/2)\}.
	\end{split}
\end{equation}

This established the ranking consistency property. \QEDclosed \\

\textbf{Proof of Theorem 2}

Following the proof of Lemma 1, note that under the event $E=\big\{\max_{j\in\mathcal{M}_*}|\hat{d}_j-d_j|\leq c_0n^{-2\tau_1}\big\}$, by Condition 3, we have $\hat{d}_j\geq \frac{1}{2}c_0n^{-\tau_1}$ holds for all $j\in\mathcal{M}_*$. Hence, we obtain $\mathcal{M}_*\subset\hat{\mathcal{M}}_{\gamma_n}$ by certain choice of $\gamma_n$, e.g., $\gamma=\frac{1}{2}c_0n^{-\tau_1}$. The result follows from a simple bound
$$Pr\big(E^c\big)\leq s_nb_1\exp\{a_n\},$$
where $a_n=-b_5n^{1-2\tau_1}-\tau_1b_3\log(b_6n)$, $b_1, b_5, b_3, b_6$ are positive constants. This completes the proof.\rightline{\QEDclosed}\\

\textbf{Proof of Corollary 1:}

Note that $\#\{j:d_j\geq\frac{1}{4}c_0n^{-\tau_1}\}\leq\frac{4}{c_0}n^{\tau_1}\sum_jd_j,$ where $\#\{\cdot\}$ denotes the size of the set $\{\cdot\}$. Hence on the set 
$$\mathcal{D}\equiv\big\{\max_{1\leq j\leq p_n}|\hat{d}_j-d_j|\leq\frac{1}{2}c_0n^{-\tau_1}\big\},$$
we have $\#\{j:\hat{d}_j\geq\frac{1}{2}c_0n^{-\tau_1}\}\leq \#\{j:d_j\geq\frac{1}{4}c_0n^{-\tau_1}\}\leq \frac{4}{c_0}n^{\tau_1}\sum_jd_j$. Therefore
$$Pr\biggl(|\hat{\mathcal{M}}|\leq \frac{4}{c_0}n^{\tau_1}\sum_jd_j\biggl)\geq P(\mathcal{D})\geq 1-p_nb_1\exp\{c_n\},$$
where $c_n=-b_8n^{2\tau_1}-b_3\tau_1\log(b_9n)$ for some constants $b_8$ and $b_9$.\\
\rightline{\QEDclosed}

\clearpage

\section{Web Appendix B}
	\textbf{Simulation Example.} 
	In this example we illustrate the transcendence of $d_j$ against $d_{1j}$, $d_{2j}$ and other correlations under various of complex nonlinear relationship between the survival outcome and the covariates. 
	
 	Let $X$ and $Z$ be univariate standard normal random variables representing the signal and the noise variable. Then, we generate the survival time by the following 5 transformation models (i) $h(T)=-0.5I(X\leq -1)-0.5I(X\geq 1)+\epsilon$, (ii) $h(T)=0.5I(X\leq -1)+\epsilon$, (iii) $h(T)=0.5I(X\geq 1)+\epsilon$, (iv) $h(T)=0.1((X/2)^3+X^2+X/3-0.5)+\epsilon$, (v) $h(T)=0.3\cos(3X+0.5)+\epsilon$, and (vi) $h(T)=0.5x+\epsilon$, where $\epsilon$ is the random noise following $N(0,0.6^2)$ and $H(t)=\log\{0.5(\exp(2t)-1)\}$. In particular, model (i) can mimic the situation that high and low levels of certain gene expression are found in non-tumor samples while tumor samples characterized with moderate expression. Censoring times are generated independently from a uniform distribution [0, u] such that the average censoring rate is around 20\%. We compute the proposed $\hat{d}$, $\hat{d}_{1}$,$\hat{d}_{2}$, and other 3 measures in the literature, SII from Li et al. (2020), correlation rank from Zhang et al. (2017), and censored rank from Song et al. (2014), for $X$ and $Z$, respectively under 100 replicates. Then the corresponding proportion that the measure for signal exceeding the measure for noise are summarized in Table \ref{tab:table1}.

	\begin{table}[h]
		\caption{Model comparison under 6 measurements.}
  \label{tab:table1}
		\begin{center}
				\scalebox{0.9}{
			\begin{tabular}{c|c|c|c|c|c|c}
				\hline
				& Model (i) & Model (ii) & Model(iii) & Model(iv) & Model(v) & Model (vi)\\
				\hline
				$\hat{d}_1$& 94\% & 98\% & 86\% & 78\% & 83\% & 100\%\\
				$\hat{d}_2$& 94\% & 82\% & 97.5\% & 94\% & 75\% & 100\%\\
				$\hat{d}$& 97\% & 95\% & 95.5\% & 93\% & 83\% & 100\%\\
				SII& 68\% & 54.5\% & 77\% & 84.5\% & 58.5\% & 100\%\\
				Correlation Rank& 28.5\% & 59\% & 61\% & 28\% & 40\% & 100\%\\
				Censored Rank& 30\% & 51\% & 66\% & 32\% & 44.5\% & 100\%\\
				\hline
			\end{tabular}
		}
		\end{center}
	\end{table}
	In general, $\hat{d}$, combining $\hat{d}_1$ and $\hat{d}_2$ has the best performance to distinguish between signal and noise with a relative high proportion of identify true signal among all 6 models and 6 measurements. Models (ii) and (iii) show $\hat{d}$ benefits from both $\hat{d}_1$ and $\hat{d}_2$, and is more robust. 
	
	To further show how the proposed $\hat{d}$ works, we investigate into flexible coefficients in models (i) and (iv) and consider the model $h(T)=c(I(X\leq -1)+I(X\geq 1))+\epsilon$ with $c=0,0.5,1,2$ and $h(T)=c((X/2)^3+X^2+X/3-0.5)+\epsilon$ with varying $c=0,0.1,0.5,1$. Then, the boxplot of $\hat{d}$ against different $c$ are drawn in Figure 1. When $c=0$, i.e., $X$ and $T$ are independent, $\hat{d}$ is very close to 0. As $c$ increases, both the spectrum and the average level of $\hat{d}$ becomes larger above from zero dramatically. 
	
	\begin{figure}[ht]
		\centering 
		\includegraphics[width=0.7\textwidth]{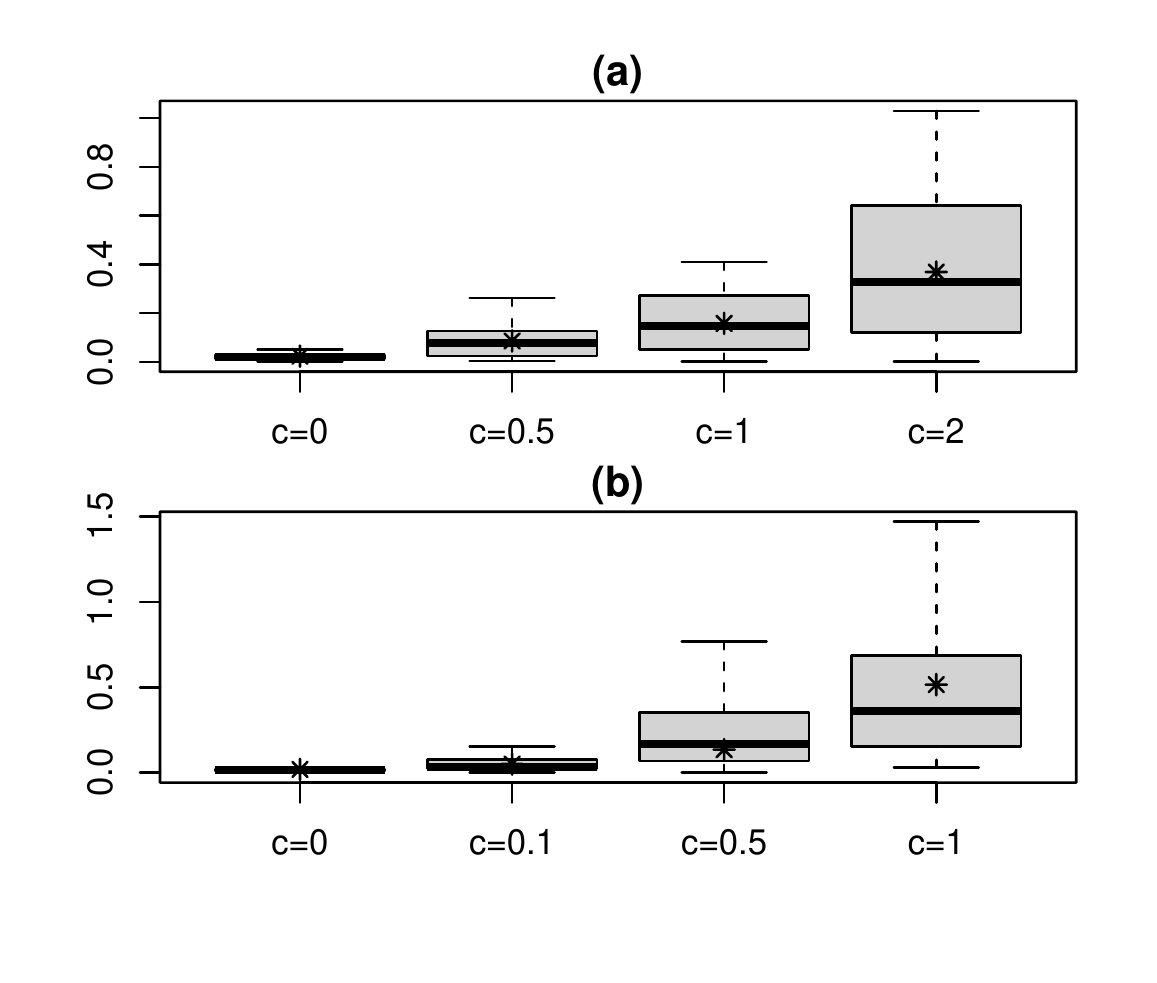} 
		\caption{(a) Boxplot of RMST difference $\hat{d}$ against $c$ of Model (i); (b) Boxplot of $\hat{d}$ against $c$ of Model (iv). The mean values are marked as stars.} 
		\label{Fig.main2} 
	\end{figure}
	
\section{Web Appendix C}
Simulation results for interval-censored data are listed in Table \ref{Tab:Table4}.
The fitted spline for the selected S1PR4 and HSD11B1 in Section 4 are shown in the following Figure \ref{fig}.

\begin{table}
		\caption{Simulation results for interval censored data in Scenario 1-4 under $p=1000$: the method proposed by Hu et al. (2020) is named as `SurvD'.}
  \label{Tab:Table4}
		\begin{tabular}{ccccccccc}
			\hline
			&	&   & Median & IQR & Pall & Median & IQR & Pall\\
			Scenario & Error  & Method & n=100  &     &      & n=200  &     &     \\
			\hline
			Scenario 1 & N(0,1) & RMST & 4 & 1 & 100\% &4&0&100\%\\
			      & & SurvD & 4 & 2.5 & 100\%&4&0&100\%\\
			& Standard Extreme & RMST & 5 & 1 & 100\%& 4&0&100\%\\
			& & SurvD & 4 & 1 & 100\% &4&0&100\%\\
			& Logistic & RMST & 4 & 0 & 100\% & 4 & 0 & 100\% \\
			& & SurvD & 4 & 0 & 100\% & 4 & 0 & 100\%\\
		    \hline
			Scenario 2 & N(0,1) & RMST & 25 & 22.5 & 45\% & 5 & 1 & 100 \%\\
			      & & SurvD & 42.5 &50.25 & 26\%& 7.5 & 4.5 & 100\%\\
			& Standard Extreme & RMST & 6 & 2 & 60\% & 5 & 1 & 100\%\\
			& & SurvD & 34 & 42.25 & 25\% & 7 & 4 & 100\%\\
			& Logistic & RMST & 26 & 26.25 & 40\% & 6 & 2.75 & 100\% \\
			& & SurvD & 47 & 51 & 12\% & 8.5 & 9.5 & 93\%\\
		    \hline
		Scenario 3 & N(0,1) & RMST & 21 & 60 & 53\% & 5 & 0 & 100\%\\
			& & SurvD & 19 & 63 & 16\%& 8 & 8 & 95\%\\
			& Standard Extreme & RMST & 23 & 170 & 50\% & 3 & 7 & 90\%\\
			& & SurvD & 54 & 172 & 18\% & 4 & 8 & 90\%\\
			& Logistic & RMST & 37 & 70 & 11\% & 36 & 123 & 54\% \\
			& & SurvD & 79 & 96 & 5\% & 92 & 202 & 19\%\\
			\hline
		    Scenario 4 & N(0,1) & RMST & 91 & 120 & 2\% & 12 & 12 & 95\%\\
			& & SurvD & 185 & 204 & 2\%& 33 & 38 & 55\%\\
			& Standard Extreme & RMST & 92.75 & 70 & 9\% & 7.5 & 9 & 96\%\\
			& & SurvD & 193.75 & 126 & 6\% & 40 & 30 &57\%\\
			& Logistic & RMST & 215 & 272 & 0\% & 21 & 29 & 75\% \\
			& & SurvD & 300 & 288 & 0\% & 37 & 58 & 50\%\\
		    \hline
		\end{tabular}
	\end{table}
 
\begin{figure}
    \centering
    \includegraphics[scale=0.6]{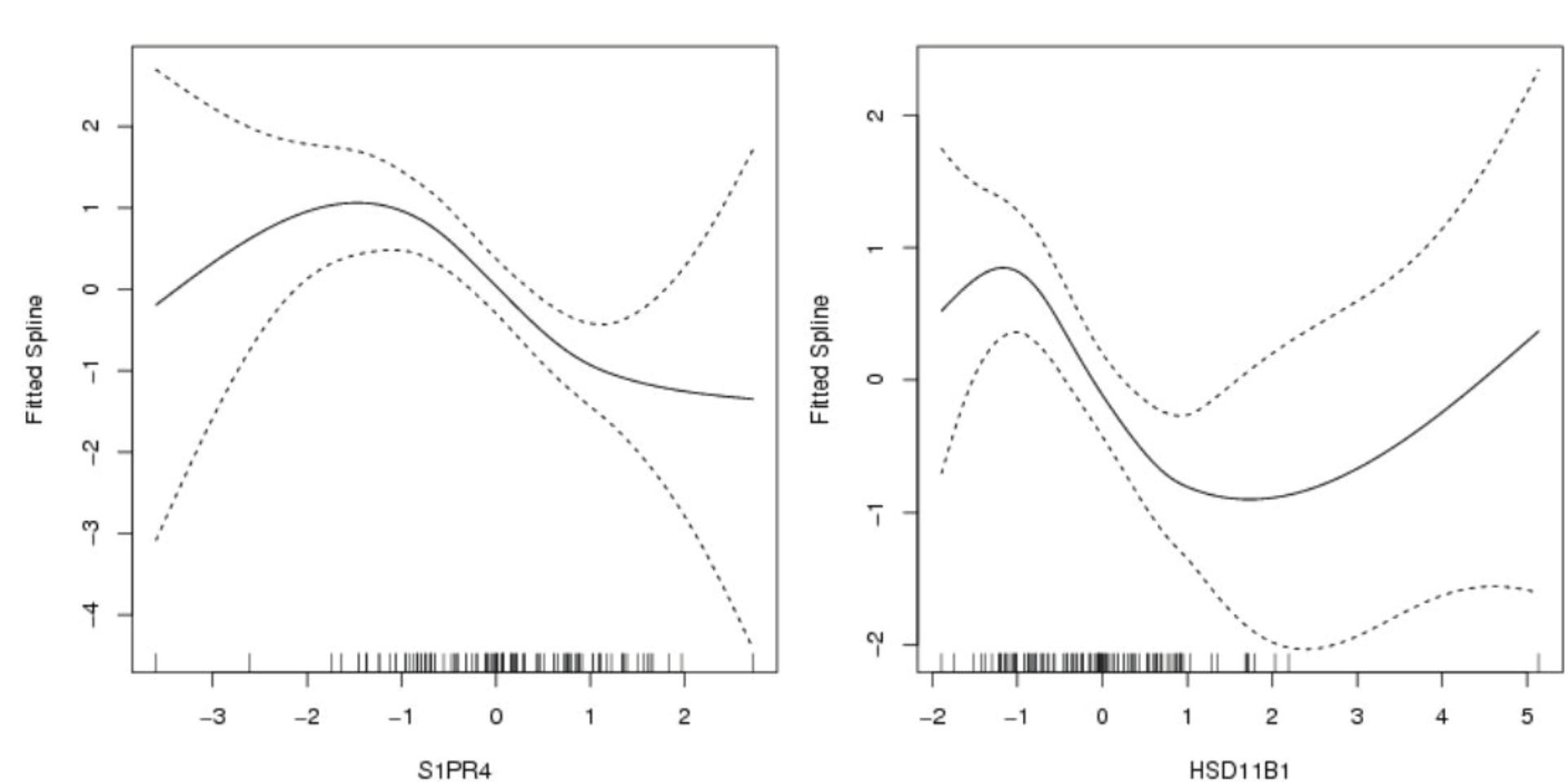}
    \caption{Fitted splines for S1PR4 and HSD11B1.}
    \label{fig}
\end{figure}